%% file: root_ExtendedVersion.tex
\documentclass[10pt]{IEEEtran}
\input{preamble.tex}

\title{\textcolor{navy}{\textsc{Dynamic Centrality Measures for Water Distribution Network Hydraulics}}}

\author{MirSaleh Bahavarnia$^\dagger$, Salma M. Elsherif$^\dagger$, and Ahmad F. Taha$^{\dagger,\star}$ 
	\thanks{$^\dagger$The authors are with the Department of Civil and Environmental Engineering, Vanderbilt University, 2201 West End Avenue, Nashville, TN 37235, USA. $^\star$Ahmad F. Taha is also affiliated with the Department of Electrical and Computer Engineering. Emails: \{mirsaleh.bahavarnia,salma.m.elsherif,ahmad.taha\}@vanderbilt.edu.}
 \thanks{This work is supported by the National Science Foundation (NSF), United States, under grants ECCS 2151571, CMMI 2152450, and CMMI 2606146.}
}

\begin{document}

\maketitle

\begin{abstract}
Water distribution networks (WDNs) are susceptible to various failures, including but not limited to human errors, cyber-attacks, and network modifications, necessitating the vulnerability analysis of WDNs. Graph-theoretic centrality measures---as a main class of centrality measures---aim to rank the network components solely based on their influence (i.e., criticality) on the WDN topology in the case of input changes, while overlooking the WDN dynamics. To overcome such a limitation, this paper uses a control-theoretic centrality measure to identify the network's most and least influential pipes in WDNs, by simultaneously incorporating the dynamics and topology of the WDN. First, given a WDN modeled by nonlinear differential-algebraic equations (NDAEs) consisting of transient flow dynamics as differential equation (DE) and conservation of water mass as algebraic equation (AE) and considering the pipe flow rates as states of the state-space (SS) representation, we extract a linearized system modeled by linear ordinary differential equations (LODEs) around the equilibrium (i.e., steady) flow rate vector. Second, treating pipe flow rates as SS nodes (states), we introduce a node centrality-based measure, namely \textit{vulnerability vector (VV)}, to rank the network pipes based on their influence on the dynamics and topology of the WDN in the case of input changes. In particular, the network's most and least influential pipes can be identified through such a centrality-based approach. This enables water engineers to understand the WDN's vulnerability better and effectively prioritize the maintenance and operational efforts on the most influential pipes within the WDN.
\end{abstract}

\begin{IEEEkeywords}
Centrality measures, edge centrality, node centrality, observability Gramian, vulnerability, water distribution networks.
\end{IEEEkeywords}

\section{Introduction and Paper Contributions}\label{sec:Intro}

\IEEEPARstart{W}{ATER} distribution networks (WDNs) are classified as one of the most critical physical infrastructure networks, as the daily life of human beings is directly intertwined with their performance. WDNs are susceptible to various damages/failures \cite{berardi2008development} such as human errors (e.g., erroneous pipe flow analysis), cyber-attacks (e.g., long-term negative effects on pipes), and network modifications (e.g., pipe aging), necessitating the vulnerability analysis of WDNs. Over the last two decades, two classes of measures have mainly been proposed to quantify the vulnerability of the networks to damage \cite{yazdani2012water}: (\textit{i}) hydraulic simulation-based measures, and (\textit{ii}) graph-theoretic centrality measures. Hydraulic simulation-based measures require the hydraulic model information of WDNs, while graph-theoretic centrality measures mainly rely on the topological structure of WDNs. On the one hand, although the former class of measures provides a comprehensive understanding of the vulnerability of WDNs, it is highly data-dependent and generally becomes computationally cumbersome, specifically for large-scale WDNs. On the other hand, the latter offers a computationally efficient yet less accurate \textit{first approximation} \cite{yazdani2012water} for the vulnerability analysis of WDNs. Therefore, these two classes of measures complement each other, and combining them provides insights that can serve as a basis for improving maintenance and operational efforts effectively.

Various hydraulic simulation-based centrality measures have been proposed in the literature. The authors in \cite{arulraj1995concept} have proposed a simple static significant index (SI) to rank the WDNs' pipes in terms of the hydraulic and physical specifications---flow rate, the Hazen-Williams roughness coefficient, length, and diameter---of each pipe. In \cite{vairavamoorthy2005pipe}, a simple hydraulic information-based pipe index (PI) is presented to rank the pipes in WDNs based on their vulnerability against damage. Through a sensitivity analysis, \cite{izquierdo2008sensitivity} has proposed a method to determine the relative importance of pipes in WDNs. By sequential node removal according to a pressure-based connectivity loss criterion, \cite{shuang2014node} has proposed a method to identify the critical nodes of WDNs. In \cite{laucelli2015vulnerability}, the consequences of pipe failure due to earthquakes are studied in terms of unsupplied demand to customers. Also, the worst pipe failure scenarios are identified by formulating a multi-objective combinatorial problem, which is solved using a multi-objective genetic algorithm. In addition, two hydraulic simulation-based centrality measures, namely \textit{flow} and \textit{leakages}, are utilized in \cite{simone2018centrality}.

The most suitable graph-theoretic centrality measures for physical infrastructure networks \cite{borgatti2005centrality}, particularly for spatially distributed systems like WDNs, are
(\textit{i}) \textit{betweenness} \cite{freeman1977set,narayanan2014little,soldi2015resilience,agathokleous2017robustness,simone2018centrality,ulusoy2018hydraulically,giustolisi2019tailoring,zarghami2019entropy,zarghami2020domain} that quantifies the significance of a node/edge in terms of its placement on the shortest paths or random-walks (alternative paths), (\textit{ii}) \textit{closeness} \cite{freeman1977set,narayanan2014little,simone2018centrality,giustolisi2019tailoring,zarghami2019entropy,zarghami2020domain} that refers to the importance of a node/edge in terms of its ability to spread the information to the other nodes/edges via minimum distance paths, and (\textit{iii}) \textit{degree} \cite{nieminen1974centrality,freeman1977set,yazdani2012water,soldi2015resilience,giustolisi2017network,simone2018centrality,giustolisi2019tailoring}, as the most intuitive centrality measure, determines the significance of a node/edge based on the number of its adjacent nodes/edges. Note that depending on the network component type, i.e., edge (e.g., pipe) or node (e.g., junction), we can use either \textit{edge} centrality measures or \textit{node} centrality measures. Tab. \ref{tab:CM} depicts the graph-theoretic centrality measure types (edge, node, or both) covered by the aforementioned research works. As Tab. \ref{tab:CM} shows, the majority of the research works have utilized the node centrality measures to rank the influence (i.e., criticality) of the network's nodes (e.g., junctions) against the failures. However, a relatively small number of research works have employed edge centrality measures to identify the network's most and least influential edges (e.g., pipes). Then, one can conclude that there is potential room for more improvements in developing edge centrality measures for WDNs to identify the network's most and least influential pipes. In addition to the three main graph-theoretic centrality measures, i.e., betweenness, closeness, and degree, several graph-theoretic centrality measures have also been proposed, some of them as modified/extended versions of the three main graph-theoretic centrality measures. Those graph-theoretic centrality measures include, but are not limited to, neighborhood nodal degree \cite{giustolisi2017network,simone2018centrality,giustolisi2019tailoring}, WFEBC (Water Flow Edge Betweenness Centrality) equipped with the modified random-walk edge betweenness centrality \cite{ulusoy2018hydraulically}, eigenvector \cite{narayanan2014little,zarghami2020domain}, Page-Rank \cite{gutierrez2013application}, HITS (Hyperlinked Induced Topic Search) \cite{gutierrez2013application}, entropy \cite{yazdani2012water,zarghami2019entropy}, demand \cite{yazdani2012water,zarghami2020domain}, Jensen-Shannon divergence (information theory) \cite{ponti2021novel}, and Wasserstein distance (optimal transport theory) \cite{ponti2021novel}. 

\begin{table}[t]
    \centering
    \caption{Graph-theoretic centrality measure types (edge, node, or both) covered by the research works in the literature.}
    \begin{tabular}{lcc}
        \toprule
        \rowcolor{tableblue} \thead{Research works} & \thead{Edge centrality} & \thead{Node centrality} \\
        \midrule
        \cite{yazdani2012water,agathokleous2017robustness,giustolisi2017network,zarghami2019entropy,zarghami2020domain} & \xmark & \checkmark \\
        \cite{soldi2015resilience,ulusoy2018hydraulically} & \checkmark & \xmark \\
        \cite{narayanan2014little,simone2018centrality,giustolisi2019tailoring} & \checkmark & \checkmark \\
        \bottomrule
    \end{tabular}
    \label{tab:CM}
\end{table} 

Although \cite{yazdani2012water,ulusoy2018hydraulically,giustolisi2019tailoring,zarghami2020domain} have enhanced the purely graph-theoretic centrality measures by taking node demands information into account, such enhanced graph-theoretic centrality measures are still far away from being a representative centrality measure to capture both topological structure and hydraulic model information systematically. To reduce such a gap, as an effective remedy, we adopt and adapt the control-theoretic centrality measure that simultaneously incorporates both model dynamics and topological structure \cite{chanekar2023gramian}. Considering the state-space (SS) representation of a network, investigating the network's response to impulse inputs injected into each \textit{SS node} (state), and introducing a node centrality measure, the authors in \cite{chanekar2023gramian} construct a \textit{vulnerability matrix (VM)} to identify the network's most and least influential SS nodes (i.e., network nodes) (Note that an SS node and a network node coincide in the SS representation utilized therein, unlike the case of our study).

As just exemplified, in the node centrality measures literature, for the case of general networks, each SS node usually represents a network node, i.e., an SS node and a network node coincide. However, in this paper, for the case of a WDN, each SS node (state) represents an edge with a pipe flow rate, implying that an SS node and a network node are distinct notions in our study. Adopting and adapting the approach proposed by \cite{chanekar2023gramian} to the context of WDNs, we consider the SS representation of a WDN, investigate the network's response to various inputs (step, pulse, and impulse inputs) injected into each SS node (i.e., pipe flow rate in the case of a WDN), and construct a node centrality-based measure, namely \textit{vulnerability vector (VV)} to identify the network's most and least influential SS nodes (i.e., pipe flow rates in the case of a WDN). We also highlight that, in terms of computational time, the computation of VV is more efficient than the computation of VM due to the modifications we have introduced compared to the approach proposed by \cite{chanekar2023gramian}. As a crucial point to clarify, we highlight that although the proposed VV specialized for a WDN is built upon a node centrality measure, it is considered an edge centrality measure as it technically identifies the network's most and least influential pipes (edges). To put the proposed VV into perspective in comparison with the graph-theoretic edge centrality measures depicted by Tab. \ref{tab:CM}, we highlight that it simultaneously incorporates both model dynamics and topological structure of the WDN, taking advantage of its control-theoretic nature. An earlier version of this work appeared in~\cite{bahavarnia2025influential}, where we studied a preliminary version of this problem by deriving the VV formulas for the shifted step and pulse inputs and identifying the most and least influential pipes of the benchmark WDNs. Compared to~\cite{bahavarnia2025influential}, the current paper includes the complete proofs of the propositions, validates the effectiveness of the VVs against the quasi-exact variations, extends the VV computation to abrupt input changes (impulse inputs), provides a graph-theoretic characterization of the insignificant components, introduces a VV-based node centrality measure, and studies the dependency of the VV-based influences on the input types, the timing parameters, and the linearization point, along with a comparison against the graph-theoretic edge betweenness centrality.

\parhead{Paper Contributions.} This paper aims to identify the network's most and least influential pipes in WDNs. The main contributions can be categorized as follows:
\begin{itemize}
    \item Given a WDN modeled by nonlinear differential-algebraic equations (NDAEs) consisting of flow dynamics as a differential equation (DE) and conservation of water mass as an algebraic equation (AE), we need to obtain a linearized system to quantify the influence of pipes via a centrality measure. To that end, considering the pipe flow rates as states of the SS representation, we extract a linearized system modeled by linear ordinary differential equations (LODEs) around the equilibrium (i.e., steady) flow rate vector, from the NDAE-modeled WDN.
    \item Treating pipe flow rates as SS nodes (states), we introduce a control-theoretic, centrality-based measure coined vulnerability vector (VV). VVs rank network pipes based on their influence on the dynamics and topology of the WDN. In particular, the network's most and least influential pipes can be identified through such a centrality-based approach. This enables water system engineers to understand the WDN's vulnerability better and effectively prioritize the maintenance and operational efforts on the most influential pipes within a WDN.
    \item The theoretical developments certify that there exist mathematical expressions for VVs in terms of the dynamics and topology of the network. Such mathematical expressions facilitate vulnerability analysis without requiring computationally expensive exact evaluation of the variations associated with the various inputs. Moreover, through extensive numerical simulations conducted on the benchmark WDNs, we investigate the effects of input types and timing parameters on the VV-based influence of the network pipes. 
    
    Interestingly, we discover a graph-theoretic interpretation for the VV-based network's insignificant pipes (i.e., pipes with VV equal to $0$). Specifically, we propose a graph-theoretic iterative procedure to characterize the subset of the VV-based network's insignificant components. Furthermore, built upon the proposed VV edge centrality measure, we suggest an intuitive node centrality measure to identify the network's most and least influential nodes that can be interpreted as the dynamic version of the static graph-theoretic centrality measure, i.e., degree. 
\end{itemize}

\begin{figure}[!t]
\centering
\begin{tikzpicture}[
  node distance=6.5mm and 4.5mm,
  box/.style={draw=navy, line width=0.5pt, rounded corners=2pt, fill=paleblue, align=center,
              font=\footnotesize, inner sep=3pt, text width=2.35cm, minimum height=9.5mm},
  lbl/.style={font=\footnotesize\bfseries, text=navy},
  arr/.style={-{Latex[length=1.6mm]}, line width=0.55pt, draw=navy}]
  \node[box] (A) {\textbf{\textcolor{navy}{WDN model}}\\ NDAE \eqref{NDAEs}: flows $q(t)$,\\ heads $h(t)$, inputs $u(t)$};
  \node[box, right=of A] (B) {\textbf{\textcolor{navy}{Linearization}}\\ equilibrium $q^{\ast}$ via \eqref{NRM},\\ LDAE \eqref{LDAEs}};
  \node[box, right=of B] (C) {\textbf{\textcolor{navy}{Reduction}}\\ null space of $C_v$,\\ LODE \eqref{LODEs}: $(F,G)$};
  \node[box, below=of B] (E) {\textbf{\textcolor{navy}{Vulnerability vector}}\\ $v_m$ in \eqref{GamForm} via\\ Propositions \ref{Propo1}--\ref{Propo3}};
  \node[box, left=of E] (D) {\textbf{\textcolor{navy}{Input changes}}\\ shifted step, pulse, impulse\\ $\mathcal{S}_1[\cdot]$, $\mathcal{S}_2[\cdot]$, $\mathcal{S}_3[\cdot]$};
  \node[box, right=of E] (F) {\textbf{\textcolor{navy}{Ranking}}\\ most and least influential\\ pipes and nodes};
  \draw[arr] (A) -- (B);
  \draw[arr] (B) -- (C);
  \draw[arr] (C.south) -- ++(0,-3.25mm) -| (E.north);
  \draw[arr] (D) -- (E);
  \draw[arr] (E) -- (F);
\end{tikzpicture}
\caption{Overview of the proposed approach. The NDAE model of the WDN is linearized around the equilibrium flow rate vector and reduced to an LODE model. The response of the LODE model to shifted step, pulse, and impulse input changes yields the vulnerability vector, which ranks the pipes and nodes of the WDN.}
\label{fig:overview}
\end{figure}
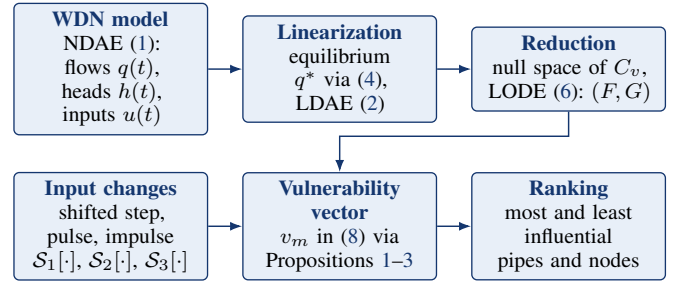

\parhead{Paper Structure.} The remainder of the paper is structured as follows: Section \ref{sec:ProFor} presents preliminaries on NDAE and LODE models of WDNs, followed by a statement of the problem to be investigated in this paper. Section \ref{VVC} contains the paper's main results on computing VV given a WDN modeled by NDAEs: Sections \ref{VVa} and \ref{VVb} are devoted to computing VV for two classes of inputs (shifted step and pulse inputs), respectively. Section \ref{newSec} is divided into three parts as follows: Section \ref{VVc} validates the effectiveness of VVs, Section \ref{Sec4B} details VV computation in the case of abrupt input changes, and Section \ref{SecD} investigates the VV-based network's insignificant components (pipes). Through extensive numerical simulations conducted on the benchmark WDNs, Section \ref{NuSim} validates the effectiveness of the centrality-based approach based on the identification quality of the network's most and least influential pipes. Finally, Section \ref{Con} expresses a few concluding remarks. Fig. \ref{fig:overview} summarizes the overall approach of the paper.

\parhead{Paper Notation.} The lowercase and uppercase letters represent the vectors and matrices, respectively. We denote the set of $n$-dimensional real-valued vectors and $m \times n$ real-valued matrices by $\mathbb{R}^n$ and $\mathbb{R}^{m \times n}$, respectively. We represent the set of positive integer numbers and the set of non-negative real numbers by $\mathbb{N}$ and $\mathbb{R}_+$, respectively. For a $J \in \mathbb{N}$, we denote $\{1,\dots,J\}$ by $\mathbb{N}_J$. For $j \in \mathbb{N}_J$, we represent the $j$-th canonical unit vector by $\mathrm{e}_j^J$. We denote the unit step function by $\mathrm{1}_{\mathbb{R}_+}(t)$ defined as $\mathrm{1}_{\mathbb{R}_+}(t) := \begin{cases}
    1 & t \ge 0,\\
    0 & t < 0,
\end{cases}$. Also, we denote the unit impulse function (also known as the Dirac delta function) by $\delta(t)$. For a real vector $x$, we represent the $i$-th element of $x$, the element-wise absolute value of $x$, and the Euclidean norm of $x$ by $x_i$, $|x|$, and $\|x\|$, respectively. Symbol $\odot$ denotes the vector element-wise product. For a real vector with non-zero elements, $x^{-1}$ denotes the element-wise inverse of $x$. For a real matrix $X$, we represent its transpose and vectorization by $X^\top$ and $\mathrm{vec}(X)$, respectively. For a real square matrix $X$, we denote its spectral abscissa, inverse (if invertible), and matrix exponential by $\mathrm{sa}(X)$, $X^{-1}$, and $e^X$, respectively. Given a real vector $x$, by $\mathrm{dg}(x)$, we mean a diagonal matrix consisting of the elements of $x$ with $x_j$ as the $jj$-th element of the diagonal matrix. We denote the $n$-dimensional zero vector, $n$-dimensional all-one vector, and $n$-dimensional identity matrix by $0_n$, $1_n$, and $I_n$, respectively. To represent the partial differentiation with respect to $x$, we utilize $\frac{\partial (.)}{\partial x}$. For a set $S$, by $\max~S$, $\sup~S$, and $\mathbf{card}(S)$, we mean the maximum element of $S$, the supremum (i.e., the least upper bound) associated with elements of $S$, and the cardinality (i.e., the number of elements) of $S$, respectively. A real symmetric matrix $X$ is called positive-definite if its eigenvalues are all positive, and we denote such positive-definiteness by $X \succ 0$. Symbol $\otimes$ represents the Kronecker product.

\section{Preliminaries and Problem Statement} \label{sec:ProFor}

This section is comprised of two main parts: \textit{(A)} preliminaries on NDAE and LODE models of WDNs, and \textit{(B)} problem statement. 

\subsection{Preliminaries on NDAE and LODE Models of WDNs}

In the current study, we adopt the NDAE-modeled WDN from \cite{masuda2019dynamical}. Such an NDAE model is categorized as a rigid water column model with slowly time-varying pipe flows \cite{islam1998modeling} and consists of two main equations: \textit{(i)} head balance (through energy balance) DE representing the transient flow dynamics, and \textit{(ii)} mass balance AE capturing the conservation of water mass. Considering the WDN quantities summarized by Tab. \ref{tab:1} and choosing $r_m(q_m)q_m|q_m|$ as the nonlinear head loss across the $m$-th pipe (as a special case of nonlinear head loss in the form of $r_m(q_m)q_m|q_m|^{\gamma-1}$ \cite{boulos2006comprehensive}), we can construct the following NDAE representation of WDNs \cite{masuda2019dynamical,bellos2018friction} (see Appendix \ref{CNDAE} for the details):
\begin{subequations} \label{NDAEs}
    \begin{align} 
    & \dot{q}(t) = B [r(q(t)) \odot q(t) \odot |q(t)|] -BC_f^\top h_f + Bu(t),\\
    & 0 = C_v q(t) + \kappa^{\mathrm{ext}},
\end{align}
\end{subequations}where $q(t)$, $h_f$, $u(t)$, and $\kappa^{\mathrm{ext}}$ denote the flow rate through the pipes, the total head associated with the nodes with fixed total heads, the input term associated with the valves on the pipes, and the external demand of water, respectively, and the detailed expressions are as follows:
\begin{align*}
    & q(t) \in \mathbb{R}^M,u(t) \in \mathbb{R}^M,r(q) = \begin{bmatrix}
        r_1(q_1) & \cdots & r_M(q_M)
    \end{bmatrix}^\top,\\
    & r_m(q_m) = \frac{8l_m}{g \pi^2 d_m^5}f_m[\rho_m(q_m)],\rho_m(q_m) = \frac{4}{\pi d_m \nu}|q_m|,\\
    & f_m(\rho_m) = \bigg(\frac{64}{\rho_m}\bigg)^{a(\rho_m)} \bigg(0.75 \ln \bigg (\frac{\rho_m}{5.37} \bigg ) \bigg)^{2(a(\rho_m)-1)b(\rho_m)} \times \\
    & \bigg(0.88 \ln \bigg (\frac{6.82 d_m}{\epsilon} \bigg ) \bigg)^{2(a(\rho_m)-1)(1-b(\rho_m))},\\
    & a(\rho_m) = \bigg (1 + \bigg (\frac{\rho_m}{2712} \bigg )^{8.4} \bigg )^{-1},\\
    & b(\rho_m) = \bigg (1 + \bigg (\frac{\epsilon \rho_m}{150d_m} \bigg )^{1.8} \bigg )^{-1}, \forall m \in \mathbb{N}_M,\\
    & B := B(C_v,D) = DC_v^\top (C_vDC_v^\top)^{-1}C_vD-D \in \mathbb{R}^{M \times M},\\
    & D := D(\iota) = \mathrm{dg}(\iota^{-1}) \in \mathbb{R}^{M \times M},\iota_m = \frac{4l_m}{g \pi d_m^2},\forall m \in \mathbb{N}_M,\\
    & C := \begin{bmatrix}
        C_v^\top & C_f^\top
    \end{bmatrix}^\top \in \mathbb{R}^{N \times M},C_v \in \mathbb{R}^{N_v \times M},\\
    & C_f \in \mathbb{R}^{N_f \times M},h(t) := \begin{bmatrix}
        h_v(t)^\top & h_f^\top
    \end{bmatrix}^\top \in \mathbb{R}^N,\\
    & h_v(t) \in \mathbb{R}^{N_v}, h_f \in \mathbb{R}^{N_f}.
\end{align*}

It is noteworthy that NDAE dynamics \eqref{NDAEs} depend on the WDN topology via the incidence matrix $C$, the pipe inertia constants of the matrix $D$, the pipe length $l$, the pipe diameter $d$, and the total head for nodes with a fixed total head $h_f$. Also, $h_v(t)$ depends on $h_f$ and can be computed as follows:
\begin{align*}
    (C_vDC_v^\top)^{-1}C_vD [r(q(t)) \odot q(t) \odot |q(t)| -C_f^\top h_f + u(t)].
\end{align*}

While water head $h(t)$ is a critical performance variable in WDNs and is interdependent with the flow rate $q(t)$ (see \eqref{TFTP} in Appendix \ref{CNDAE}), the flow-driven dynamic formulation adopted in this paper enables pipe-level states that directly capture connectivity, transport, and operational continuity. This perspective facilitates a broader set of insights and operational considerations by ranking pipes according to their dynamic and topological influence, thereby indicating which pipes would have the greatest operational impact when disruptions such as leaks, breaks, or flow interruptions occur.

We highlight that the proposed VV is developed for a specific NDAE dynamics. However, the main dynamic control-theoretic centrality-based idea can be adopted for any general WDN describable by the standard state-space representation.

\begin{table}[t]
    \centering
    \caption{Summary of WDN quantities.}
    \begin{tabular}{cl}
        \toprule
        \rowcolor{tableblue} \thead{Quantity} & \thead{Description} \\
        \midrule
        $N$ & Number of nodes (e.g., junctions) \\
        $M$ & Number of edges (pipes) \\
        $t$ & Time ($\mathrm{s}$) \\
        $q_m(t)$ & Time-dependent flow rate through the $m$-th pipe \\
        $h_i(t)$ & Total head at the $i$-th node \\
        $r_m(q_m)$ & Pipe coefficient of the $m$-th pipe \\
        $u_m(t)$ & Input term coming from a valve on the $m$-th pipe \\
        $\iota_m$ & Pipe inertia constant of the $m$-th pipe \\
        $f_m(\rho_m)$ & Darcy-Weisbach friction factor for the $m$-th pipe \\
        $l_m$ & Length of the $m$-th pipe \\
        $d_m$ & Diameter of the $m$-th pipe \\
        $g$ & Gravitational acceleration: $9.80665~\mathrm{m/s^2}$ \\
        $\rho_m(q_m)$ & Reynolds number for the $m$-th pipe \\
        $\nu$ & Kinematic viscosity of water: $1.007 \times 10^{-6}~\mathrm{m^2/s}$ \\
        $\epsilon$ & Roughness coefficient of cast iron: $2.591 \times 10^{-4}~\mathrm{m}$ \\
        $h_f$ & Total head for nodes with fixed total heads \\
        $h_v(t)$ & Total head for nodes with varying total heads \\
        $N_f$ & Number of nodes with fixed total heads \\
        $N_v$ & Number of nodes with varying total heads \\
        $D$ & Edge weights diagonal matrix \\
        $C$ & Incidence matrix \\
        $\kappa^{{ext}}_i$ & External demand of water at the $i$-th node \\
        \bottomrule
    \end{tabular}
    \label{tab:1}
\end{table}

In the sequel, we convert the NDAE model to an LDAE model as we need to obtain a linearized system to quantify the influence of pipes via a centrality measure.

\subsubsection{Conversion of the NDAE model to an LDAE model}

Linearizing NDAE dynamics \eqref{NDAEs} around the equilibrium flow rate vector and the equilibrium input vector, namely $(q^{\ast},u^{\ast})$, we get the following LDAE representation:
\begin{subequations} \label{LDAEs}
    \begin{align}
    & \dot{\bar{q}}(t) = B \mathrm{dg}[(2r^{\ast} + r'^{\ast} \odot q^{\ast}) \odot |q^{\ast}|]\bar{q}(t) + B \bar{u}(t), \label{LDE}\\
    & 0 = C_v \bar{q}(t),\label{LAE}
\end{align}
\end{subequations}
where the equilibrium deviation quantities are as follows: 
\begin{align*}
    & \bar{q}(t) := q(t)-q^{\ast}, \bar{u}(t) := u(t)-u^{\ast}, q^{\ast} = \begin{bmatrix}
        q_1^{\ast} & \cdots & q_M^{\ast}
    \end{bmatrix}^\top,\\
    & r^{\ast} = \begin{bmatrix}
        r_1^{\ast} & \cdots & r_M^{\ast}
    \end{bmatrix}^\top, r'^{\ast} = \begin{bmatrix}
        r'^{\ast}_1 & \cdots & r'^{\ast}_M
    \end{bmatrix}^\top,\\
    & r_m^{\ast} := r_m(q_m^{\ast}), r'_m(q_m) := \frac{dr_m(q_m)}{dq_m},r'^{\ast}_m := r'_m(q_m^{\ast}),\\
    & \forall m \in \mathbb{N}_M.
\end{align*}
Utilizing the chain rule in calculus, we get
\begin{align*}
    \frac{dr_m(q_m)}{dq_m} & = \frac{8l_m}{g \pi^2 d_m^5} \frac{d f_m[\rho_m(q_m)]}{dq_m}\bigg|_{q_m}\\ 
    & = \frac{8l_m}{g \pi^2 d_m^5} \frac{d f_m(\rho_m)}{d \rho_m}\bigg|_{\rho_m(q_m)} \frac{d \rho_m(q_m)}{dq_m}\bigg|_{q_m},
\end{align*}
that can be used to obtain a closed-form expression for $r'^{\ast}$ as the ones for $f_m(\rho_m)$ and $\rho_m(q_m)$ already exist.

Now, we aim to compute the equilibrium flow rate vector $q^{\ast}$. According to \eqref{TFTP} and \eqref{CWM} in Appendix \ref{CNDAE}, and by definition of the equilibrium, the following equations:\begin{subequations} \label{EqEqs}
    \begin{align} 
    & 0 = C_v^\top h_v^{\ast} + C_f^\top h_f - [r(q^{\ast}) \odot q^{\ast} \odot |q^{\ast}|] - u^{\ast},\\
    & 0 = C_v q^{\ast} + \kappa^{\mathrm{ext}},
\end{align}
\end{subequations}are satisfied. Setting $u^{\ast} = 0$, for given $h_f$ and $\kappa^{\mathrm{ext}}$, we iteratively solve \eqref{EqEqs} for $(q^{\ast},h_v^{\ast})$ by utilizing the Newton-Raphson method detailed by \cite{masuda2019dynamical} via the following updating equation:
\begin{align} \label{NRM}
    \begin{bmatrix}
        q^{(n+1)} \\ h_v^{(n+1)}
    \end{bmatrix} &= \begin{bmatrix}
        q^{(n)} \\ h_v^{(n)}
    \end{bmatrix} - [\Upsilon^{(n)}]^{-1} \notag \\
    & \hspace*{-1.6em} \times \begin{bmatrix}
        C_v^\top h_v^{(n)} + C_f^\top h_f - [r(q^{(n)}) \odot q^{(n)} \odot |q^{(n)}|] \\
        C_v q^{(n)} + \kappa^{\mathrm{ext}}
    \end{bmatrix},
\end{align}
where the Jacobian matrix at the $n$-th iteration $\Upsilon^{(n)}$ is computed as follows:
\begin{align*}
    \Upsilon^{(n)} &= \begin{bmatrix}
        -\mathrm{dg}[(2r^{(n)} + r'^{(n)} \odot q^{(n)}) \odot |q^{(n)}|] & C_v^\top \\
        C_v & 0
    \end{bmatrix},
\end{align*}
where the superscript $^{(k)}$ refers to the corresponding values at the $k$-th iteration for all $k \in \mathbb{N}$. The stopping criterion for the Newton-Raphson method is considered as follows:
\begin{align*}
    \|q^{(n+1)}-q^{(n)}\|^2 + \|h_v^{(n+1)}-h_v^{(n)}\|^2 \le \varepsilon_{\mathrm{tol}},
\end{align*}
where $\varepsilon_{\mathrm{tol}}$ denotes the error tolerance.

In the sequel, we convert the LDAE model to an LODE model.

\subsubsection{Conversion of the LDAE model to an LODE model}
To convert the LDAE model to an LODE model, we merge AE \eqref{LAE} with DE \eqref{LDE}. As AE \eqref{LAE} shows, $\bar{q}(t)$ belongs to the null space of $C_v$. Since the null space of $C_v$ is equal to the null space of $E := C_v^\top C_v$ \cite{horn1985matrix}, we can equivalently replace $0 = C_v \bar{q}(t)$ by $0 = E \bar{q}(t)$. According to the fact that $C_v$ is a full-row rank matrix and $E$ is a symmetric singular matrix, we have the following eigenvalue decomposition for $E$: 
\begin{align} \label{EVD}
    & E = \begin{bmatrix}
        U_1 & U_2
    \end{bmatrix}\mathrm{dg}(\begin{bmatrix}
        \lambda & 0_{M-N_v}
    \end{bmatrix})\begin{bmatrix}
        U_1 & U_2
    \end{bmatrix}^\top,
\end{align}
where $U_1 \in \mathbb{R}^{M \times N_v}$ and $U_2 \in \mathbb{R}^{M \times (M-N_v)}$ denote the eigenvectors associated with the eigenvalues $\lambda \in \mathbb{R}^{N_v}$ and $0_{M-N_v}$, respectively. We denote $M-N_v$ by $K$ in the sequel for notational simplicity.

According to the eigenvalue decomposition of $E$ represented by \eqref{EVD} and noting that $E = U_1 \mathrm{dg}(\lambda) U_1^\top$ and $U_1^\top U_2 = 0$ hold, we can consider $\bar{q}(t)$ as $\bar{q}(t) = U_2 \tilde{q}(t)$. Hydraulically, $\tilde{q}(t)$ can be interpreted as net imbalance flows that are actively working to equalize pressure differences across different regions of the network. This means that the new state variables act as decoupled, virtual fluid circuits. Substituting $\bar{q}(t) = U_2 \tilde{q}(t)$ in DE \eqref{LDE}, pre-multiplying both sides by $U_2^\top$, and noting that $U_2^\top U_2 = I_K$ holds, we get the following LODE representation:
\begin{subequations} \label{LODEs}
\begin{empheq}[box=\widefbox]{align} 
    & \dot{\tilde{q}}(t) = F \tilde{q}(t) + G \bar{u}(t),\\
    & F := U_2^\top B \mathrm{dg}[(2r^{\ast} + r'^{\ast} \odot q^{\ast}) \odot |q^{\ast}|]U_2 \in \mathbb{R}^{K \times K},\label{LODEsb}\\ 
    & G := U_2^\top B \in \mathbb{R}^{K \times M}. \label{LODEsc}
\end{empheq}    
\end{subequations}Remarkably, the state matrix $F$ is stable (Hurwitz), i.e., its eigenvalues all lie on the open left half plane \cite{masuda2019dynamical}. We also highlight that the equilibrium flow rate vector $q^{\ast}$ depends on the external demand $\kappa^{\mathrm{ext}}$, $C$, $D$, $l$, $d$, and $h_f$ leading to the dependency of LODE dynamics \eqref{LODEs} on the external demand $\kappa^{\mathrm{ext}}$, the WDN topology $C$, the pipe inertia constants of the matrix $D$, the pipe length $l$, the pipe diameter $d$, and the total head for nodes with a fixed total head $h_f$. The solution of LODE dynamics \eqref{LODEs} takes the following form \cite{chen1984linear}:
\begin{align} \label{Sol}
    & \tilde{q}(t) = e^{Ft}\tilde{q}(0) + \int_0^t e^{F(t-\tau)}G \bar{u}(\tau) d\tau.
\end{align}

The input change $\bar{u}(t)$ is injected into the system via virtual valves for mathematically computing the influence of the pipes and ranking them to identify the most and least influential pipes. Changing $\bar{u}(t)$ mathematically simulates an operator or actuator physically turning a valve wheel. A sudden step-change or rapid ramp down in the input represents a rapid closure, which generates a pressure surge or water hammer wave.

\subsection{Problem Statement}

Here, we present the main problem to be investigated in this paper as follows:

\begin{problembox}
\begin{mypbm}
    Given a WDN modeled by NDAE dynamics \eqref{NDAEs}, introduce and compute a centrality measure to rank the network pipes based on their influence (i.e., criticality) on the dynamics and topology of the WDN in the case of input changes.
\end{mypbm}
\end{problembox}

To assess the vulnerability of the pipes of a WDN in the case of input changes (implemented through valves, see the definition of input term $u_m(t)$ in Tab. \ref{tab:1}), treating pipe flow rates as SS nodes (states), we introduce a node centrality-based measure, VV. On the chosen input types, we remark that in control systems, step and pulse functions are standard test signals used to analyze the dynamic system behavior. A step function represents a sudden, sustained change in input. A pulse function represents a short-duration, high-intensity input used to analyze transient responses. The limit of a pulse as its duration approaches zero, with the area under the pulse kept constant, is an impulse that represents the abrupt changes in inputs. The step response is the integral of the impulse response, meaning a step provides information on how the system sustains energy, while an impulse (pulse) shows the immediate, sharp disturbance \cite{ogata2010modern}. In the case of WDNs, the application-based considerations imply that it is preferable to consider step and pulse responses instead of impulse responses as the WDN dynamics slowly change. Then, in this paper, we choose step and pulse functions as the main inputs in our theoretical derivations regarding the computation of VV. However, to investigate VV computation in the case of abrupt input changes, we include the case of an impulse function as input. Such abrupt input changes potentially cause failures. The hydraulic interpretation of the considered different input types is as follows:
\begin{itemize}
    \item Step input can hydraulically be interpreted as a sudden valve opening or a sudden load application. Practical applications include emergency stops, relief valve cracking, or rapid directional changes.
    \item Pulse input can hydraulically be interpreted as a brief, timed valve actuation or a temporary load engagement. Practical applications include inching/jogging cylinders, fast injection molding, or digital hydraulics.
    \item Impulse input can hydraulically be interpreted as a sudden, momentary pressure spike or water hammer. Practical applications include component fatigue testing or rapid switching in pilot lines.
\end{itemize}

Throughout this paper, we aim to find comprehensive answers to the following questions:

\begin{itemize}[\IEEEsetlabelwidth{Q5.}]
    \item[\qlab{Q1.}] Is it possible to incorporate both the dynamics and topology of a WDN to identify its most and least influential pipes?
    \item[\qlab{Q2.}] Can one derive mathematical expressions for VV in terms of dynamics and topology of the network without evaluating the exact variations associated with the various inputs?
    \item[\qlab{Q3.}] Is the VV-based influence of the network pipes dependent on the input types and their timing parameters?
    \item[\qlab{Q4.}] Is there any graph-theoretic interpretation for the VV-based network's insignificant components (pipes and nodes)?
    \item[\qlab{Q5.}] Can one induce a centrality measure from VV to additionally identify the network's most and least influential nodes in WDNs?
\end{itemize}

\section{VV Computation} \label{VVC}

We compute the vulnerability vector VV for each pipe in the network. Then, we denote VV by $v \in \mathbb{R}^{M}$. For WDNs, considering the step and pulse functions as inputs, we define the $m$-th element of $v$, i.e., $v_{m}$ for $m \in \mathbb{N}_M$ as follows:
\begin{align} \label{GamForm}
    & v_{m} := \mathcal{S} \bigg [ \frac{\partial \tilde{q}(t;m,\alpha)}{\partial \alpha} \bigg ],~\forall m \in \mathbb{N}_M,
\end{align}
where $\mathcal{S}[\cdot]$ represents a computational operator and $\tilde{q}(t;m,\alpha)$ denotes the state $\tilde{q}(t)$ of LODE dynamics \eqref{LODEs} with the shifted step input $\bar{u}(t) = \alpha \mathrm{1}_{\mathbb{R}_+}(t-\mu) \mathrm{e}_m^M$ or the shifted pulse input $\bar{u}(t) = \alpha [\mathrm{1}_{\mathbb{R}_+}(t-\mu)-\mathrm{1}_{\mathbb{R}_+}(t-\mu-\eta)] \mathrm{e}_m^M$ with $\alpha$ (magnitude parameter: the input flow rate magnitude), $\mu$ (shifting parameter (also known as delay parameter): the time instant at which the step signal $\mathrm{1}_{\mathbb{R}_+}(t-\mu)$ transitions from $0$ to $1$), and $\eta$ (pulse length parameter: the time duration of applying the pulse signal) as parameters incorporating the input flow rate magnitude, the time instant of applying the input change, and the time duration of applying the pulse input change, respectively. Since a pulse input is essentially a combination of a positive step input followed by a delayed negative step input, the pulse length parameter $\eta$ can also be interpreted as a delay parameter. Considering $v_{m}$ defined by \eqref{GamForm}, the influence of the $m$-th pipe can be captured via $v_m$ and the network's most and least influential pipes can be determined as $m^{\mathrm{Most~Influential}} := \argmax_{m \in \mathbb{N}_M} v_m$ and $m^{\mathrm{Least~Influential}} := \argmin_{m \in \mathbb{N}_M} v_m$, respectively.

We need different types of inputs to test how input changes to the system lead to significant changes in the pipe flow rates. To that end, we need to have specific control inputs, and we need to parameterize them with parameter $\alpha$. We need to see how different $\alpha$-dependent control inputs impact the evolution of the dynamics of \eqref{LODEs}. Accordingly, we need to choose an appropriate computational operator $\mathcal{S}[\cdot]$ to obtain a well-posed vulnerability vector VV in terms of informativeness. Depending on the form of state $\tilde{q}(t;m,\alpha)$, the well-posedness of vulnerability vector VV translates into the finiteness of the outcome of $\mathcal{S} \big [ \frac{\partial \tilde{q}(t;m,\alpha)}{\partial \alpha} \big ]$ in \eqref{GamForm}. Such well-posedness acts as a basis to choose the diverse computational operators $\mathcal{S}_1[\cdot]$, $\mathcal{S}_2[\cdot]$, and $\mathcal{S}_3[\cdot]$ in the following. 

In the sequel, we first compute VV for two classes of inputs: \textit{(i)} shifted step input in Section \ref{VVa}, and \textit{(ii)} shifted pulse input in Section \ref{VVb}.

\subsection{VV Computation for the Shifted Step Input} \label{VVa}

First, we compute $\tilde{q}(t;m,\alpha)$ and $\frac{\partial  \tilde{q}(t;m,\alpha)}{\partial \alpha}$ in the sequel. Solving the dynamics \eqref{LODEs} for $\tilde{q}(t;m,\alpha)$ with the shifted step input $\bar{u}(t) = \alpha \mathrm{1}_{\mathbb{R}_+}(t-\mu) \mathrm{e}_m^M$ (by utilizing \eqref{Sol}), we get 
\begin{align*}
    & \tilde{q}(t;m,\alpha) = \begin{cases}
        e^{Ft}\tilde{q}(0), & 0 \le t < \mu\\
        e^{Ft}\tilde{q}(0) + \alpha (e^{F(t-\mu)}-I_K)p_m^K, & \mu \le t
    \end{cases},
\end{align*}
where $p_m^K := F^{-1}G \mathrm{e}_m^M \in \mathbb{R}^{K}$. Then, we have
\begin{align} \label{pSSI}
    & \frac{\partial \tilde{q}(t;m,\alpha)}{\partial \alpha} = \begin{cases}
        0_K, & 0 \le t < \mu\\
        (e^{F(t-\mu)}-I_K)p_m^K, & \mu \le t
    \end{cases}.
\end{align}

We need to appropriately choose $\mathcal{S}[\cdot]$ such that $v_{m}$ in \eqref{GamForm} takes finite values. Otherwise, it cannot quantify the influence of each pipe on the WDN dynamics. For the shifted step input, a potential candidate for $\mathcal{S}[\cdot]$ can be the following root mean square (RMS)-based candidate:
\begin{align} \label{Sch}
    & \mathcal{S}_1[x(t);T] := \sqrt{\frac{1}{T}\int_0^T \|x(t)\|^2 dt}.
\end{align}
where $T$ denotes the time horizon length. Although $\underset{t \in [0,T]}{\sup} \|x(t)\|$ can also be considered as a potential candidate for $\mathcal{S}[\cdot]$, it is computationally inefficient as its computation requires solving a non-convex optimization problem. It is noteworthy that $\mathcal{S}_1[x(t);T] \le \underset{t \in [0,T]}{\sup} \|x(t)\|$ holds. The following proposition presents a mathematical expression for VV, i.e., $v$ for the shifted step input $\bar{u}(t) = \alpha \mathrm{1}_{\mathbb{R}_+}(t-\mu) \mathrm{e}_m^M$.

\begin{myprs} \label{Propo1}
    Considering the dynamics \eqref{LODEs}, for the shifted step input $\bar{u}(t) = \alpha \mathrm{1}_{\mathbb{R}_+}(t-\mu) \mathrm{e}_m^M$ and the computational operator $\mathcal{S}_1[x(t);T]$ defined by \eqref{Sch}, the vulnerability vector can be computed via
    \begin{empheq}[box=\widefbox]{equation}
    \begin{aligned}
        & v_m = \sqrt{Y_{mm}},~\forall m \in \mathbb{N}_M,\\
        & Y := \frac{1}{T}G^\top (F^{-1})^\top \bigg(W_{T-\mu} \\
        & \quad -(e^{F^\top(T-\mu)}-I_K)(F^{-1})^\top \\
        & \quad -F^{-1}(e^{F(T-\mu)}-I_K) + (T-\mu)I_K \bigg )F^{-1}G,\\
        & F^\top W_{T-\mu} + W_{T-\mu}F + I_K \\
        & \quad -e^{F^\top(T-\mu)}e^{F(T-\mu)} = 0.
    \end{aligned}
    \label{g1}
    \end{empheq}
    where $W_{T-\mu} \in \mathbb{R}^{K \times K}$ denotes the unique positive-definite solution of the Lyapunov equation \eqref{g1} for which we have
    \begin{align*}
    W_{T-\mu} =&~ \int_0^{\infty} e^{F^\top t}(I_K -e^{F^\top(T-\mu)}e^{F(T-\mu)})e^{F t} dt,\\
    \mathrm{vec}(W_{T-\mu}) =&~ (I_K \otimes F^\top + F^\top \otimes I_K)^{-1}\\ 
    & ~\times \mathrm{vec}(e^{F^\top(T-\mu)}e^{F(T-\mu)}-I_K).
    \end{align*}
\end{myprs}
For the proof, see Appendix \ref{PF1}. It is remarkable that for $v_m$ in \eqref{g1}, we have 
\begin{align*}
    & \lim_{T \to \infty} v_m = \sqrt{\mathrm{e}_m^{M \top} G^\top (F^{-1})^\top F^{-1} G \mathrm{e}_m^M},
\end{align*}
in the final value sense. Note that $W_{T-\mu}$ can be interpreted as an observability Gramian according to \eqref{g1}. Also, we highlight that $W_{T-\mu}$ can efficiently be computed by solving the Lyapunov equation \eqref{g1} via off-the-shelf Lyapunov equation solvers. Then, VV computation can efficiently be done by the formula \eqref{g1} presented in Proposition \ref{Propo1}.

\subsection{VV Computation for the Shifted Pulse Input} \label{VVb}

Similarly, we first compute $\tilde{q}(t;m,\alpha)$ and $\frac{\partial \tilde{q}(t;m,\alpha)}{\partial \alpha}$ in the sequel. Solving the dynamics \eqref{LODEs} for $\tilde{q}(t;m,\alpha)$ with the shifted pulse input $\bar{u}(t) = \alpha[\mathrm{1}_{\mathbb{R}_+}(t-\mu)-\mathrm{1}_{\mathbb{R}_+}(t-\mu-\eta)] \mathrm{e}_m^M$ (by utilizing \eqref{Sol}), we get
\begin{align*}
    & \tilde{q}(t;m,\alpha) = \begin{cases}
        e^{Ft}\tilde{q}(0), & 0 \le t < \mu\\
        e^{Ft}\tilde{q}(0) + \alpha (e^{F(t-\mu)}-I_K)p_m^K, & \mu \le t < \xi\\
        e^{Ft}\tilde{q}(0) + \alpha (e^{F\eta}-I_K)p_m^K, & \xi \le t
    \end{cases},
\end{align*}
where $\xi = \mu + \eta$. We then have
\begin{align} \label{pSPI}
    & \frac{\partial \tilde{q}(t;m,\alpha)}{\partial \alpha} = \begin{cases}
        0_K, & 0 \le t < \mu\\
        (e^{F(t-\mu)}-I_K)p_m^K, & \mu \le t < \xi\\
        (e^{F\eta}-I_K)p_m^K, & \xi \le t
    \end{cases}.
\end{align}

For the shifted pulse input, in addition to $\mathcal{S}_1[x(t);T]$ defined by \eqref{Sch}, another potential candidate for $\mathcal{S}[\cdot]$ can be
\begin{align} \label{Schh}
    & \mathcal{S}_2[x(t)] := \max \Bigg \{\sqrt{\frac{1}{\eta}\int_{\mu}^{\xi} \|x(t)\|^2dt},\|x(\xi)\| \Bigg \},
\end{align}
which takes the maximum of the RMSs over the two distinctive time intervals. Likewise, although $\underset{t \in [0,T]}{\sup} \|x(t)\|$ can also be considered as a potential candidate for $\mathcal{S}[\cdot]$, it is computationally inefficient as its computation requires solving a non-convex optimization problem. It is noteworthy that $\mathcal{S}_1[x(t);T] \le \underset{t \in [0,T]}{\sup} \|x(t)\|$ and $\mathcal{S}_2[x(t)] \le \underset{t \in [0,\infty)}{\sup} \|x(t)\|$ hold. In the sequel, Propositions \ref{Propo2} and \ref{Propo3} express mathematical expressions for VV, i.e., $v$ for the shifted pulse input $\bar{u}(t) = \alpha[\mathrm{1}_{\mathbb{R}_+}(t-\mu)-\mathrm{1}_{\mathbb{R}_+}(t-\mu-\eta)] \mathrm{e}_m^M$.

\begin{myprs} \label{Propo2}
    Considering the dynamics \eqref{LODEs}, for the shifted pulse input $\bar{u}(t) = \alpha[\mathrm{1}_{\mathbb{R}_+}(t-\mu)-\mathrm{1}_{\mathbb{R}_+}(t-\mu-\eta)] \mathrm{e}_m^M$ and the computational operator $\mathcal{S}_1[x(t);T]$ defined by \eqref{Sch}, the vulnerability vector can be computed via
    \begin{empheq}[box=\widefbox]{equation}
    \begin{aligned}
        & v_m = \sqrt{Y_{mm}},~\forall m \in \mathbb{N}_M,\\
        & Y := \frac{1}{T} G^\top (F^{-1})^\top \bigg(W_\eta -(e^{F^\top \eta}-I_K)(F^{-1})^\top \\
        & -F^{-1}(e^{F\eta}-I_K) + \eta I_K \\
        & + (T-\mu-\eta)(e^{F^\top \eta}-I_K)(e^{F \eta}-I_K) \bigg )F^{-1}G,\\
        & F^\top W_\eta + W_\eta F + I_K -e^{F^\top \eta}e^{F\eta} = 0.
    \end{aligned}
    \label{g2}
    \end{empheq}
    where $W_{\eta} \in \mathbb{R}^{K \times K}$ denotes the unique positive-definite solution of the Lyapunov equation \eqref{g2} for which we have
    \begin{align*}
    W_{\eta} =&~ \int_0^{\infty} e^{F^\top t}(I_K -e^{F^\top \eta}e^{F \eta})e^{F t} dt,\\
    \mathrm{vec}(W_{\eta}) =&~ (I_K \otimes F^\top + F^\top \otimes I_K)^{-1}\\ 
    & ~\times \mathrm{vec}(e^{F^\top \eta}e^{F \eta}-I_K).
    \end{align*}
\end{myprs}
For the proof, see Appendix \ref{PF2}. Remarkably, for $v_m$ in \eqref{g2}, we get the following expression for $\underset{T \to \infty}{\lim} v_m$:
\begin{align*}
     & \sqrt{\mathrm{e}_m^{M \top} G^\top (F^{-1})^\top (e^{F^\top \eta}-I_K)(e^{F \eta}-I_K) F^{-1} G \mathrm{e}_m^M}
\end{align*}
in the final value sense. Also, note that $W_{\eta}$ can similarly be interpreted as an observability Gramian according to \eqref{g2}.

\begin{myprs} \label{Propo3}
    For the shifted pulse input $\bar{u}(t) = \alpha[\mathrm{1}_{\mathbb{R}_+}(t-\mu)-\mathrm{1}_{\mathbb{R}_+}(t-\mu-\eta)] \mathrm{e}_m^M$ and the computational operator $\mathcal{S}_2[x(t)]$ defined by \eqref{Schh}, the vulnerability vector can be computed via
    \begin{empheq}[box=\widefbox]{equation}
    \begin{aligned}
        & v_m = \max \{\sqrt{Y_{mm}},\sqrt{Z_{mm}}\},~\forall m \in \mathbb{N}_M,\\
        & Y := \frac{1}{\eta} G^\top (F^{-1})^\top \bigg(W_\eta -(e^{F^\top \eta}-I_K)(F^{-1})^\top \\
        & -F^{-1}(e^{F\eta}-I_K) + \eta I_K \bigg ) F^{-1}G,\\
        & Z := G^\top (F^{-1})^\top(e^{F^\top \eta}-I_K)(e^{F \eta}-I_K)F^{-1}G.
    \end{aligned}
    \label{g3}
    \end{empheq}
    where $W_{\eta} \in \mathbb{R}^{K \times K}$ denotes the unique positive-definite solution of the Lyapunov equation \eqref{g2}.
\end{myprs}
For the proof, see Appendix \ref{PF3}.

\section{VV Effectiveness, Abrupt Input Changes, and Insignificant Components} \label{newSec}

This section is divided into three main parts: \textit{(A)} effectiveness of VVs is validated by comparing their values to the quasi-exact variations associated with the inputs, \textit{(B)} VV computation in the case of abrupt input changes is investigated, and \textit{(C)} built upon the formulas for VVs, the network's insignificant components (pipes) are identified.

We highlight that since NDAE dynamics \eqref{NDAEs} do not admit a mathematical expression as a solution (unlike LODE dynamics \eqref{LODEs} that admit \eqref{Sol} as a solution), computing the exact variations associated with the inputs is computationally cumbersome. Therefore, we compute the exact variations associated with the inputs for LODE dynamics \eqref{LODEs} instead. Since doing this causes an approximation due to the linearization of NDAE dynamics \eqref{NDAEs}, we utilize the term \textit{quasi-exact} variations associated with the inputs with respect to NDAE dynamics \eqref{NDAEs}.

Solving the system of NDAEs in \eqref{NDAEs} for exact variations in WDNs is computationally expensive because: \textit{(i)} WDN dynamics couple differential equations with complex nonlinear algebraic equations. Exact evaluation requires running iterative numerical solvers (like the Global Gradient Algorithm (GGA) developed by \cite{61052.61053}) at every single time step, and \textit{(ii)} evaluating vulnerabilities across a WDN requires simulating numerous potential input variations, e.g., different consumer demand patterns, pipe bursts, or cyber-physical pumps/valves overrides.

The VV-based vulnerability analysis is an offline planning and risk-assessment tool, not a real-time control system. Yet, computational efficiency remains vital for WDN management because, first, the offline security assessments must screen thousands of potential threat vectors across extended, week-long planning horizons, and second, the analysis must deliver rapid, actionable data for day-ahead scheduling or post-incident mapping.

The main advantages of using the VV-based approach are as follows: \textit{(i)} it bypasses the need to fully re-solve the massive, nonlinear algebraic network equations iteratively for every minor input modification, \textit{(ii)} it captures the essential transient behavior and hydraulic sensitivities accurately without the full simulation overhead, and \textit{(iii)} it scales efficiently to large-scale municipal water networks with thousands of nodes and pipes.

\subsection{Effectiveness Validation of VVs} \label{VVc}

To validate the effectiveness of VVs proposed by Propositions \ref{Propo1}--\ref{Propo3}, we define the following quantity:
    \begin{align} \label{EVs}
        & J_m(\alpha,T) := \underset{t \in [0,T]}{\sup} \|\tilde{q}(t;m,\alpha)\|,
    \end{align}
    to evaluate the quasi-exact variation associated with the shifted step input $\bar{u}(t) = \alpha \mathrm{1}_{\mathbb{R}_+}(t-\mu) \mathrm{e}_m^M$ and the shifted pulse input $\bar{u}(t) = \alpha [\mathrm{1}_{\mathbb{R}_+}(t-\mu)-\mathrm{1}_{\mathbb{R}_+}(t-\mu-\eta)] \mathrm{e}_m^M$ where $\tilde{q}(t;m,\alpha)$ can be computed via the closed-form expressions presented by Sections \ref{VVa} and \ref{VVb}. Hydraulically, $J_m(\alpha,T)$ can be interpreted as the peak value for the variations of the net imbalance flows associated with the input changes injected to the network via virtual valves. Note that unlike VVs proposed by Propositions \ref{Propo1}--\ref{Propo3}, $J_m(\alpha,T)$ depends on $\alpha$ and $\tilde{q}(0)$. We emphasize that evaluating $J_m(\alpha,T)$ is computationally challenging as it necessitates solving a non-convex optimization problem. Specifically, for the case of large-scale WDNs (i.e., larger values of $M$), since we need to repeat such computation $M$ times, the situation becomes even more complicated. Moreover, by evaluating $\tilde{q}(t;m,\alpha)$ at the interval boundaries, it can be verified that for all $m \in \mathbb{N}_M$, $J_m(\alpha,T)$ is lower bounded by the following bounds:
    \begin{align*}
        J_m(\alpha,T) \ge &~ \max \{\| \tilde{q}(0)\|,\|e^{F \mu} \tilde{q}(0)\|,\\ 
        &~ \|e^{F T}\tilde{q}(0) + \alpha (e^{F(T-\mu)}-I_K)F^{-1}G \mathrm{e}_m^M\| \},\\
        J_m(\alpha,T) \ge &~ \max \{\| \tilde{q}(0)\|,\|e^{F \mu} \tilde{q}(0)\|,\\ 
        &~ \| e^{F(\mu+\eta)}\tilde{q}(0) + \alpha (e^{F \eta} - I_K)F^{-1}G \mathrm{e}_m^M\|,\\ 
        &~ \| e^{F T}\tilde{q}(0) + \alpha (e^{F \eta} - I_K)F^{-1}G \mathrm{e}_m^M\| \},
    \end{align*}
for the shifted step input $\bar{u}(t) = \alpha \mathrm{1}_{\mathbb{R}_+}(t-\mu) \mathrm{e}_m^M$ and the shifted pulse input $\bar{u}(t) = \alpha [\mathrm{1}_{\mathbb{R}_+}(t-\mu)-\mathrm{1}_{\mathbb{R}_+}(t-\mu-\eta)] \mathrm{e}_m^M$, respectively. Particularly, for the final value scenario, i.e., $T \to \infty$, the corresponding lower bounds on $\underset{T \to \infty}{\lim} J_m(\alpha,T)$ respectively boil down to the following forms:
\begin{align*}
    & \max \{\| \tilde{q}(0)\|,\|e^{F \mu} \tilde{q}(0)\|, |\alpha| \|F^{-1}G \mathrm{e}_m^M\| \},\\
    & \max \{\| \tilde{q}(0)\|,\|e^{F \mu} \tilde{q}(0)\|, |\alpha| \| (e^{F \eta} - I_K)F^{-1}G \mathrm{e}_m^M\|,\\ 
        & \| e^{F(\mu+\eta)}\tilde{q}(0) + \alpha (e^{F \eta} - I_K)F^{-1}G \mathrm{e}_m^M\|\},
\end{align*}
for all $m \in \mathbb{N}_M$. 

We highlight the fact that the VV metric is a first-order type approach to rank the pipes of the WDN in terms of their criticality in the network. In other words, the VV metric is a computationally efficient method to bypass the computational difficulty originating from the evaluation of the quasi-exact variation defined by \eqref{EVs}. Intuitively, the VV metric effectively takes into account the first-order approximations to provide a network operator with practical insights regarding the contribution level of each pipe to the variation $J_m(\alpha,T)$ caused by the input change (i.e., disturbance input) at that pipe. It is noteworthy that the term ``insignificant" is interpreted in the sense of the VV metric, not the quasi-exact variation $J_m(\alpha,T)$ or the exact variation associated with the NDAE model expressed by \eqref{NDAEs}.

In a nutshell, since evaluating $J_m(\alpha,T)$ is computationally cumbersome due to the aforementioned limiting factors, specifically for the case of large-scale WDNs (i.e., larger values of $M$), the proposed vulnerability vector VV provides us with an insightful computational tool to effectively quantify the influence of the pipes in WDNs.

We highlight VV computation in the case of abrupt input changes, potentially causing failures in the sequel.

\subsection{Abrupt Input Changes Scenario} \label{Sec4B}

To investigate VV computation in the case of abrupt input changes, we consider the shifted impulse input $\bar{u}(t) = \alpha \delta(t-\mu) \mathrm{e}_m^M$ with an abrupt input change occurring at time instant $\mu$. Solving the dynamics \eqref{LODEs} for $\tilde{q}(t;m,\alpha)$ with the shifted impulse input $\bar{u}(t) = \alpha \delta(t-\mu) \mathrm{e}_m^M$ (by utilizing \eqref{Sol}), we get
    \begin{align*}
    & \tilde{q}(t;m,\alpha) = \begin{cases}
        e^{Ft}\tilde{q}(0), & 0 \le t < \mu\\
        e^{Ft}\tilde{q}(0) + \alpha e^{F(t-\mu)} G \mathrm{e}_m^M, & \mu \le t
    \end{cases},
\end{align*}
with
\begin{align*}
    & \frac{\partial \tilde{q}(t;m,\alpha)}{\partial \alpha} = \begin{cases}
        0_K, & 0 \le t < \mu\\
        e^{F(t-\mu)} G \mathrm{e}_m^M, & \mu \le t
    \end{cases}.
\end{align*}
For the shifted impulse input $\bar{u}(t) = \alpha \delta(t-\mu) \mathrm{e}_m^M$, choosing the energy-based computational operator $\mathcal{S}_3[x(t);T] := \int_0^T \|x(t)\|^2 dt$, it can be verified that the vulnerability vector can be computed via $v_m = \mathrm{e}_m^{M \top} G^\top W_{T-\mu} G \mathrm{e}_m^M$ for all $m \in \mathbb{N}_M$ where $W_{T-\mu} \in \mathbb{R}^{K \times K}$ denotes the unique positive-definite solution of the Lyapunov equation \eqref{g1}. Also, in the final value sense, we have $\underset{T \to \infty}{\lim} v_m = \mathrm{e}_m^{M \top} G^\top W_{\infty} G \mathrm{e}_m^M$ where $W_{\infty} \in \mathbb{R}^{K \times K}$ denotes the unique positive-definite solution of the Lyapunov equation $F^\top W_{\infty} + W_{\infty} F + I_K = 0$. Also, utilizing \eqref{EVs}, we can evaluate the quasi-exact variation associated with the shifted impulse input $\bar{u}(t) = \alpha \delta(t-\mu) \mathrm{e}_m^M$ via the corresponding closed-form expressions. Moreover, by evaluating $\tilde{q}(t;m,\alpha)$ at the interval boundaries, it can be verified that for all $m \in \mathbb{N}_M$, $J_m(\alpha,T)$ is lower bounded by the following bounds:
\begin{align*}
    J_m(\alpha,T) \ge &~ \max \{ \|\tilde{q}(0)\|, \|e^{F\mu} \tilde{q}(0)\|, \|e^{F\mu} \tilde{q}(0) + \alpha G \mathrm{e}_m^M\|,\\ 
    &~ \|e^{F T} \tilde{q}(0) + \alpha e^{F(T-\mu)} G \mathrm{e}_m^M \| \},
\end{align*}
for the shifted impulse input $\bar{u}(t) = \alpha \delta(t-\mu) \mathrm{e}_m^M$. Particularly, for the final value scenario, i.e., $T \to \infty$, the corresponding lower bound on $\underset{T \to \infty}{\lim} J_m(\alpha,T)$ boils down to the $\max \{ \|\tilde{q}(0)\|,\|e^{F\mu} \tilde{q}(0)\|, \|e^{F\mu} \tilde{q}(0) + \alpha G \mathrm{e}_m^M\| \}$ for all $m \in \mathbb{N}_M$. We highlight that in the case of the equilibrium initial state ($\bar{q}(0)=0$), if $F + F^\top$ is Hurwitz, then $J_m(\alpha,T) = |\alpha| \|G \mathrm{e}_m^M\|$ holds because for $G \mathrm{e}_m^M \neq 0$ and $t \ge \mu$, we have
\begin{align*}
    & \frac{d \|\tilde{q}(t;m,\alpha)\|}{dt} = \frac{d \sqrt{\alpha \mathrm{e}_m^{M \top} G^\top e^{F^\top(t-\mu)}\alpha e^{F(t-\mu)}G \mathrm{e}_m^{M}}}{dt}\\
    & = |\alpha| \frac{\mathrm{e}_m^{M \top} G^\top e^{F^\top(t-\mu)}(F+F^\top)e^{F(t-\mu)}G \mathrm{e}_m^{M}}{2 \sqrt{\mathrm{e}_m^{M \top} G^\top e^{F^\top(t-\mu)}e^{F(t-\mu)}G \mathrm{e}_m^{M}}} < 0,
\end{align*}
implying that $\|\tilde{q}(t;m,\alpha)\|$ is a decreasing function of $t$ for $t \ge \mu$ and it takes its supremum value at $t = \mu$, i.e., $J_m(\alpha,T) = |\alpha| \|G \mathrm{e}_m^{M}\|$.

Note that for $\tilde{q}(0) \neq 0$, if $F+F^\top$ is Hurwitz, then $\| \tilde{q}(0)\| > \|e^{F \mu} \tilde{q}(0)\|$ holds, because for $\tilde{q}(0) \neq 0$ and $t \in [0,\mu)$, we have
\begin{align*}
    & \frac{d \|\tilde{q}(t;m,\alpha)\|}{dt} = \frac{d \|e^{Ft} \tilde{q}(0)\|}{dt} = \frac{d \sqrt{\tilde{q}(0)^\top e^{F^\top t}e^{Ft}\tilde{q}(0)}}{dt}\\ 
    & = \frac{\tilde{q}(0)^\top e^{F^\top t} (F+F^\top) e^{F t}\tilde{q}(0)}{2\sqrt{\tilde{q}(0)^\top e^{F^\top t}e^{Ft}\tilde{q}(0)}} < 0,
\end{align*}
implying that $\|\tilde{q}(t;m,\alpha)\|$ is a decreasing function of $t$ for $t \in [0,\mu)$ and it takes its supremum value at $t = 0$, i.e., $\| \tilde{q}(0)\| > \|e^{F \mu} \tilde{q}(0)\|$ leading to the simplification of the corresponding lower bounds on $J_m(\alpha,T)$ by elimination of the term $\|e^{F \mu} \tilde{q}(0)\|$ inside the $\max$ operator.

\subsection{VV-based Network's Insignificant Components (Pipes)} \label{SecD}

This section sheds light on a graph-theoretic interpretation of the VV-based network's insignificant components (pipes).
 
According to VV formulas presented by Propositions \ref{Propo1}--\ref{Propo3}, we realize that $B \mathrm{e}_m^M = 0$ is a sufficient condition to ensure that $v_m = 0$, i.e., the $m$-th pipe is insignificant in the sense of VV. The following two-part proposition enables us to identify a subset of insignificant components.

\begin{myprs} \label{Propo4}
    (i) If the $i$-th node with a varying total head is degree-$1$ (i.e., leaf), then the $m_i$-th pipe satisfies $B \mathrm{e}_m^M = 0$ for $m = m_i$, (ii) If the $i$-th and $j$-th nodes with varying total heads are degree-$1$ and degree-$2$, respectively, and they are adjacent, then in addition to the $m_i$-th pipe, the $m_j$-th pipe also satisfies $B \mathrm{e}_m^M = 0$ for $m = m_j$.
\end{myprs}

For the proof, see Appendix \ref{PF4}. Denoting the underlying graph associated with the WDN by $\mathcal{G}$ and centering around results of Proposition \ref{Propo4}, the rows of the corresponding $C_v$ can be considered to find potential solutions to $B \mathrm{e}_m^M = 0$ via the graph-theoretic iterative approach summarized by Procedure \ref{proc:unICC}.

Note that in Procedure \ref{proc:unICC}, for any $m \in \underline{\mathcal{I}}^{\mathrm{IC}}$, $B \mathrm{e}_m^M = 0$ is satisfied and consequently $v_m = 0$ is satisfied, i.e., the $m$-th pipe is insignificant. Moreover, the set of solutions to $B \mathrm{e}_m^M = 0$ obtained by Procedure \ref{proc:unICC} is a subset of the main set of solutions to $B \mathrm{e}_m^M = 0$ (and consequently the main set of solutions to $v_m = 0$). Therefore, it provides a lower bound $\underline{n}^{\mathrm{IC}} := \mathbf{card}(\underline{\mathcal{I}}^{\mathrm{IC}})$ for the total number of insignificant components (pipes) $n^{\mathrm{IC}}$.

Regarding the topological and hydraulic interpretation of the insignificant components, we also highlight that matrix $B = B(C_v,D) = D C_v^\top (C_v D C_v^\top)^{-1}C_vD-D$ depends on $C_v$ and $D$ where $C_v$ denotes the submatrix of the incidence matrix $C := \begin{bmatrix}
        C_v^\top & C_f^\top
    \end{bmatrix}^\top \in \mathbb{R}^{N \times M}$ associated with the nodes with varying total heads and $D := D(\iota) = \mathrm{dg}(\iota^{-1}) \in \mathbb{R}^{M \times M},\iota_m = \frac{4l_m}{g \pi d_m^2},\forall m \in \mathbb{N}_M$ denotes the edge weights diagonal matrix dependent on the hydraulic information parameters $l_m$ (length of the $m$-th pipe) and $d_m$ (diameter of the $m$-th pipe).

    Built upon the formulas for VVs, the network's most and least influential nodes can be identified as detailed in Appendix \ref{SecEConvertedtoAppendix}.

{\setlength{\floatsep}{5pt}
\begin{algorithm}[t]
\caption{Subset of Insignificant Components}\label{proc:unICC}
\DontPrintSemicolon

\KwInput{$\mathcal{G}$}

Set the initial graph $\mathcal{G}_0 = \mathcal{G}$.

Set $n \leftarrow 1$.

\While{at least one edge with a degree-$1$ node with a varying total head exists in $\mathcal{G}_{n-1}$}{

Consider $\mathcal{G}_{n-1}$ and simultaneously remove all edges with a degree-$1$ node with a varying total head in one shot to get $\mathcal{G}_{n}$.

Set $n \leftarrow n+1$.}

Construct the graph-theoretic subset of insignificant components $\underline{\mathcal{I}}^{\mathrm{IC}}$ by including the index $m$ associated with any removed edge.

\KwOutput{$\underline{\mathcal{I}}^{\mathrm{IC}}$}
\end{algorithm}
}

\section{Numerical Simulations} \label{NuSim}

In this section, we compute VV for various scenarios employing the $80$ synthesized WDNs \cite{meng2018topological} utilized by \cite{masuda2019dynamical} for the computation of the local stability index defined as the absolute value of the maximum real part of the eigenvalues of $B \mathrm{dg}[(2r^{\ast} + r'^{\ast} \odot q^{\ast}) \odot |q^{\ast}|]$. Such a local stability index is considered a resilience metric for the WDN. The higher the local stability index is, the more resilient the WDN is. The information $(N,N_f,M)$ of the $80$ benchmark WDNs is presented in Tab. \ref{tab:2}. All the simulations have been performed in MATLAB R2024a. The diameter of most of the pipes is set to $400~\mathrm{mm}$. Also, a few pipes with larger diameters exist (e.g., with a diameter of $900~\mathrm{mm}$). The software \cite{de2014hydrogen} automatically generates the diameter of the pipes. We set the total head of all the $N_f$ reservoirs in each benchmark WDN to $65~\mathrm{m}$, i.e., $h_f = 65 \times 1_{N_f}~\mathrm{m}$. In addition to these $N_f$ nodes, some nodes, e.g., pure junctions connecting the pipes in each benchmark WDN, have zero water demand, i.e., $\kappa_i^{\mathrm{ext}} = 0~\mathrm{m^3/s}$. We also set $\varepsilon = 10^{-2}$. To compute $W_{T-\mu}$ and $W_{\eta}$ in \eqref{g1} and \eqref{g2}, respectively, we utilize the MATLAB built-in Lyapunov equation solver, namely $\texttt{lyap}(.,.)$, as follows:
\begin{align*}
    W_{T-\mu} & = \texttt{lyap}\Big(F^\top,I_K -e^{F^\top(T-\mu)}e^{F(T-\mu)}\Big),\\
    W_{\eta} & = \texttt{lyap}\Big(F^\top,I_K -e^{F^\top \eta}e^{F \eta}\Big).
\end{align*}
Procedure \ref{proc:VVC} summarizes VV computation for the shifted step and pulse inputs along with their corresponding computational operator $\mathcal{S}[\cdot]$. By sorting the elements of $v$ obtained from Procedure \ref{proc:VVC}, we can identify the network's most and least influential pipes.

\begin{table}[t]
    \centering
    \caption{Information $(N,N_f,M)$ of the $80$ benchmark WDNs.}
    \footnotesize\setlength{\tabcolsep}{3.5pt}
    \begin{tabular}{cc@{\hspace{0.8em}}cc}
        \toprule
        \rowcolor{tableblue} \thead{Number of WDNs} & \thead{$(N,N_f,M)$} & \thead{Number of WDNs} & \thead{$(N,N_f,M)$} \\
        \midrule
        $10$ & $(102,2,110)$ & $9$ & $(306,6,395)$\\
        $11$ & $(102,2,130)$ & $5$ & $(404,4,443)$\\
        $10$ & $(204,4,223)$ & $5$ & $(404,4,523)$\\
        $10$ & $(204,4,263)$ & $5$ & $(406,6,525)$\\
        $9$ & $(306,6,335)$ & $5$ & $(408,8,447)$\\
        & & $1$ & $(506,6,554)$\\
        \bottomrule
    \end{tabular}
    \label{tab:2}
\end{table}

{\setlength{\floatsep}{5pt}
\begin{algorithm}[t]
\caption{VV Computation}\label{proc:VVC}
\DontPrintSemicolon
\KwInput{$l$, $d$, $C$, $h_f$, $\kappa^{\mathrm{ext}}$, $\varepsilon_{\mathrm{tol}}$, $\bar{u}(t)$, $\mathcal{S}[\cdot]$, $\mu$, $\eta$, $T$}

Setting $u^{\ast} = 0$, for given $h_f$ and $\kappa^{\mathrm{ext}}$, iteratively solve \eqref{EqEqs} for $q^{\ast}$ via the Newton-Raphson method \eqref{NRM} and compute $r^{\ast}$ and $r'^{\ast}$ accordingly.

Construct $D$ via $D := D(\iota) = \mathrm{dg}(\iota^{-1}),\iota_m = \frac{4l_m}{g \pi d_m^2},\forall m \in \mathbb{N}_M$.

Construct $B$ via $B := B(C_v,D) = DC_v^\top (C_vDC_v^\top)^{-1}C_vD-D$.

Extract $U_2$ via the eigenvalue decomposition \eqref{EVD} of $E := C_v^\top C_v$.

Construct $F$ and $G$ via \eqref{LODEsb} and \eqref{LODEsc}, respectively.
    
    \For{$m = 1$ \KwTo $M$}{\If{$\bar{u}(t) = \alpha \mathrm{1}_{\mathbb{R}_+}(t-\mu) \mathrm{e}_m^M$}{Compute $v_m$ via \eqref{g1}.}
    
    \If{$\bar{u}(t) = \alpha[\mathrm{1}_{\mathbb{R}_+}(t-\mu)-\mathrm{1}_{\mathbb{R}_+}(t-\mu-\eta)] \mathrm{e}_m^M$}{\If{$\mathcal{S}[\cdot]=\mathcal{S}_1[x(t);T]$ \textup{defined by} \eqref{Sch}}{Compute $v_m$ via \eqref{g2}.} 
    
    \If{$\mathcal{S}[\cdot]=\mathcal{S}_2[x(t)]$ \textup{defined by} \eqref{Schh}}{Compute $v_m$ via \eqref{g3}.}
    
    }
    }

Construct $v$ via $v = \begin{bmatrix}
    v_1 & \cdots & v_M
\end{bmatrix}^\top$.

\KwOutput{$v$}
\end{algorithm}
}

\subsection{Network's Most and Least Influential Components}

\begin{figure}[!t]
    \centering
    \includegraphics[width=\columnwidth]{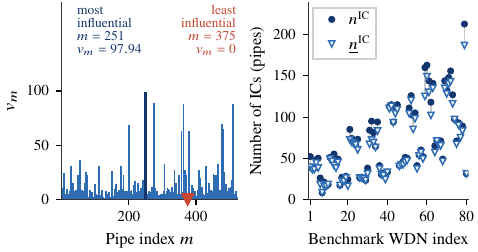}
\caption{[Left]: Distribution of the VV values across all pipes for the $66$th benchmark WDN. Each bar corresponds to a pipe, indexed arbitrarily. The highlighted bar corresponds to the most influential pipe and the marker on the horizontal axis corresponds to one representative least influential (particularly, insignificant) pipe. [Right]: total number of insignificant components (pipes) $n^{\mathrm{IC}}$ and the lower bound $\underline{n}^{\mathrm{IC}}$ presented by Section \ref{SecD} across all $80$ benchmark WDNs.}
    \label{fig:1}
\end{figure}

\begin{figure}[!t]
    \centering
    \includegraphics[width=\columnwidth]{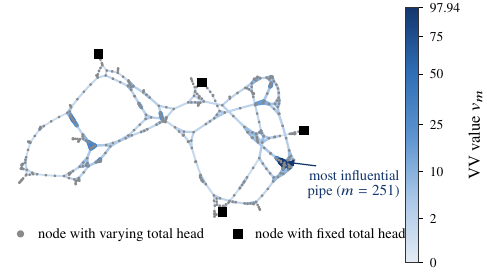}
\caption{Heatmap of the VV values across all pipes for the $66$th benchmark WDN with gray dots and black squares representing the nodes with varying and fixed total heads, respectively, and color-coded line segments representing the pipes. The intensity of the blue coloration and the width of the line segments increase with the influence of the pipe (see the color bar). The most influential pipe is indicated by the arrow.}
    \label{fig:2}
\end{figure}

First, we set $T_{\max} = \frac{-5}{\mathrm{sa}(F)}$ as a maximum upper bound for the time horizon length $T$. Then, we choose $T = c_T \times T_{\max}$, $\mu = c_{\mu} \times T$, and $\eta = c_{\eta} \times (T-\mu)$, respectively where $0 < c_T \le 1$, $0 < c_\mu < 1$, and $0 < c_\eta < 1$. For the $66$th benchmark WDN (randomly selected among the 80 benchmark WDNs), we have $\mathrm{sa}(F) = -9.1479 \times 10^{-4}$ and $T_{\max} = \frac{-5}{\mathrm{sa}(F)} = 5465.7353$. Setting $(c_T,c_\mu,c_\eta) = (0.9,0.25,0.2)$, i.e., $(T,\mu,\eta) = (0.9T_{\max},0.225T_{\max},0.135T_{\max}) = (4919.1618,1229.7904,737.8743)$ and running Procedure \ref{proc:VVC}, we realize that the most influential pipe is the $251$st pipe connecting the $252$nd node and the $364$th node. Also, the least influential pipe (in this particular case, identical to the insignificant pipe) is the $375$th pipe connecting the $376$th node and the $390$th node. Fig. \ref{fig:1} [Left] depicts the elements of VV $v_m$ versus pipe index $m$ along with the highlighted network's most and least influential pipes for the $66$th benchmark WDN. Fig. \ref{fig:1} [Right] visualizes the total number of insignificant components (pipes) $n^{\mathrm{IC}}$ and the lower bound $\underline{n}^{\mathrm{IC}}$ presented by Section \ref{SecD} for all $80$ benchmark WDNs. Fig. \ref{fig:2} visualizes the heatmap of the VV values across all pipes for the $66$th benchmark WDN.

\begin{figure}[!t]
    \centering
    \includegraphics[width=\columnwidth]{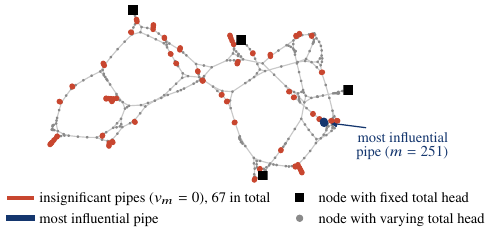}
\caption{Insignificant pipes and the most influential pipe for the $66$th benchmark WDN with gray dots and black squares representing the nodes with varying and fixed total heads, respectively. The $67$ insignificant pipes and their end nodes are shown in red, and the most influential pipe is shown in blue.}
    \label{fig:GL}
\end{figure}

\begin{figure}[!t]
    \centering
    \includegraphics[width=\columnwidth]{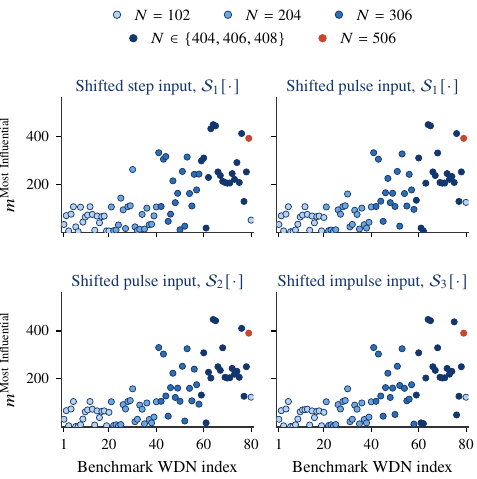}
\caption{VV-based most influential pipe metric $m^{\mathrm{Most~Influential}}$ for all $80$ benchmark WDNs. [Top Left]: shifted step input with $\mathcal{S}_1[\cdot]$, [Top Right]: shifted pulse input with $\mathcal{S}_1[\cdot]$, [Bottom Left]: shifted pulse input with $\mathcal{S}_2[\cdot]$, [Bottom Right]: shifted impulse input with $\mathcal{S}_3[\cdot]$.}
    \label{fig:MInds}
\end{figure}

We highlight that in Fig. \ref{fig:1} [Left] and Fig. \ref{fig:2}, since the least influential pipe is also an insignificant component in the sense of VV, i.e., $v_{375} = 0$ holds, it could be non-unique, which is the case for the $66$th benchmark WDN according to the iterative procedure presented by Section \ref{SecD} as $1 < \underline{n}^{\mathrm{IC}} = 66 \le n^{\mathrm{IC}} = 67$ holds. Due to illustration limitations, we have illustrated only one of the least influential pipes in Fig. \ref{fig:1} [Left] as a representative one. To overcome such illustration limitations, Fig. \ref{fig:GL} illustrates those $67$ insignificant components (pipes) and the most influential pipe overlaid on the $66$th benchmark WDN's graphical layout. We visualize the VV-based most influential pipe metric $m^{\mathrm{Most~Influential}}$ for all $80$ benchmark WDNs in Fig. \ref{fig:MInds}.

Similar to the case of the pipes, the network's most and least influential nodes can be identified, as exemplified by additional numerical simulations detailed in Appendix \ref{SecEConvertedtoAppendix}.

\subsection{Parametric Dependency of the VV-based Influences}

In this section, we visualize the dependency of the VV-based influences on three parameters $T$, $\mu$, and $\eta$, respectively.

Note that we have $\mu = c_\mu c_T T_{\max}$ and $\eta = c_\eta(1-c_\mu)c_T T_{\max}$. For fixed values of $(\mu,\eta)$, the varying values of $T$ can be chosen from the following intervals (depending on the input type):
\begin{align*}
    & (c_\mu c_T T_{\max},T_{\max}],~((c_\mu + c_\eta(1-c_\mu))c_T T_{\max},T_{\max}].
\end{align*}
For fixed values of $(T,\eta)$, the varying values of $\mu$ can be chosen from the following intervals (depending on the input type):
\begin{align*}
    & (0,c_T T_{\max}),~(0,(1-c_\eta(1-c_\mu))c_T T_{\max}).
\end{align*}
For fixed values of $(T,\mu)$, the varying values of $\eta$ can be chosen from the following interval:
\begin{align*}
    (0,(1-c_\mu)c_T T_{\max}).
\end{align*}

To enhance readability, we have moved all the figures of this section to Appendix \ref{Suppendix}.

\subsubsection{Dependency of the VV-based influences on the time horizon length} \label{DVV1} Considering $(c_T,c_\mu,c_\eta) = (0.9,0.25,0.2)$ and for fixed values of $(\mu,\eta) = (0.225 T_{\max},0.135 T_{\max}) = (1229.7904,737.8743)$, we compute the VV-based influences for different values of $T$. To that end, we consider $100$ equidistant values for $T \in (0.225 T_{\max},T_{\max}] = (1229.7904,5465.7353]$ and $T \in (0.36 T_{\max},T_{\max}] = (1967.6647,5465.7353]$ (depending on the input type). Fig. \ref{fig:3} [Top] illustrates the dependency of the VV-based influences on the time horizon length parameter $T$. As Fig. \ref{fig:3} [Top] suggests, the least influential pipe is not affected by varying $T$. Similarly, the most influential pipe is not sensitive to varying $T$ in the case of shifted pulse input (potentially due to the finite duration and amplitude of the non-zero portion of the input). However, for smaller values of $T$, the most influential pipe can change in the cases of shifted step and impulse inputs.

\subsubsection{Dependency of the VV-based influences on the shifting parameter} \label{DVV2} Considering $(c_T,c_\mu,c_\eta) = (0.9,0.25,0.2)$ and for fixed values of $(T,\eta) = (0.9T_{\max},0.135T_{\max}) = (4919.1618,737.8743)$, we compute the VV-based influences for different values of $\mu$. To that end, we consider $100$ equidistant values for $\mu \in (0,0.9T_{\max}) = (0,4919.1618)$ and $\mu \in (0,0.765T_{\max}) = (0,4181.2875)$ (depending on the input type). Fig. \ref{fig:3} [Middle] illustrates the dependency of the VV-based influences on the shifting parameter $\mu$. As Fig. \ref{fig:3} [Middle] suggests, the least influential pipe is not affected by varying $\mu$. Similarly, the most influential pipe is not sensitive to varying $\mu$ in the case of shifted pulse input (potentially due to the finite duration and amplitude of the non-zero portion of the input). However, for larger values of $\mu$, the most influential pipe can change in the cases of shifted step and impulse inputs.

\subsubsection{Dependency of the VV-based influences on the pulse length parameter} \label{DVV3} Considering $(c_T,c_\mu,c_\eta) = (0.9,0.25,0.2)$ and for fixed values of $(T,\mu) = (0.9T_{\max},0.225T_{\max}) = (4919.1618,1229.7904)$, we compute the VV-based influences for different values of $\eta$. To that end, we consider $100$ equidistant values for $\eta \in (0,0.675T_{\max}) = (0,3689.3713)$. Fig. \ref{fig:3} [Bottom] illustrates the dependency of the VV-based influences on the pulse length parameter $\eta$. As Fig. \ref{fig:3} [Bottom] suggests, the least influential pipe is not affected by varying $\eta$. However, the most influential pipe can change for smaller values of $\eta$.

It is noteworthy that changing the benchmark WDN can potentially affect the parametric dependency of the VV-based influences in Sections \ref{DVV1}--\ref{DVV3}. We highlight that the sensitivity associated with the input type or timing is not strongly noticeable for the $66$th benchmark WDN. However, it could be due to the choice of the benchmark WDN, as we have observed that some other benchmark WDNs have higher sensitivity levels. For instance, in the case of the $62$nd benchmark WDN, we obtain Fig. \ref{figg} associated with the dependency of the VV-based influences on three parameters $(T,\mu,\eta)$ under different parameter selections. As Fig. \ref{figg} represents, in the case of the $62$nd benchmark WDN---unlike the case of the $66$th benchmark WDN in Fig. \ref{fig:3}---the most influential index could be highly sensitive with respect to three timing parameters $(T,\mu,\eta)$, e.g., see Fig. \ref{figg} [Top] that depicts the trend of multiple changes in $m^{\mathrm{Most~Influential}}: 14 \to 5 \to 228 \to 433$ for the varying values of the time horizon length parameter $T$ for the shifted step input with $\mathcal{S}_1[\cdot]$.

\subsubsection{Dependency on the initialization point}In the case of linearized dynamics, the effectiveness of the applied dynamic centrality measures---in terms of accurate identification of the influential components---is \textit{theoretically} tied to the accuracy of the equilibrium state computation. One could expect a completely different ranking of the most and least influential pipes in the network. Fortunately, the derived metrics and methods seem to be somewhat robust to changes in the initialization point. To illustrate such an observation in the case studies, we identify the most influential component of the $66$th benchmark WDN for two scenarios: \textit{(i)} linearized dynamics around an unperturbed equilibrium state (i.e., previous results in earlier sections), and \textit{(ii)} linearized dynamics around $50$ randomly perturbed equilibrium states with $35\%$ and $40\%$ perturbations. Fig. \ref{figUP} [Top Left, Top Right] visualizes the identified most influential components. As Fig. \ref{figUP} [Top Left, Top Right] depicts, for even a large perturbation level on the equilibrium state, and among most perturbation scenarios, the most influential component in the network is still correctly identified. In some cases, this is not true. It seems apparent that the higher the perturbation level, the more likely the misidentification of the most influential component---a limitation of this work. We next shed light on the range of validity of the linearization point vis-\`{a}-vis the identification of the most influential pipe.

We here seek to quantify the sensitivity of the identification of the most influential component index with respect to the perturbation percentage. Specifically, for any of the equidistant perturbation percentages ranging from $0$ to $50$ (i.e., step size of $5$), we identify the most influential component of the $66$th benchmark WDN for linearized dynamics around $50$ randomly perturbed equilibrium states. We repeat the whole process $10$ times (i.e., $10$ replicates) and report the averaged values of the \textit{invariance percentage} versus the random perturbation percentage. By invariance percentage, we mean the percentage of the $50$ randomly perturbed equilibrium states for which the most influential component index is correctly identified as identical to the case of the unperturbed scenario. Fig. \ref{figUP} [Bottom] depicts the average invariance percentage versus random perturbation percentage for the $66$th benchmark WDN. As Fig. \ref{figUP} [Bottom] corroborates, for up to $15\%$ of random perturbation on the equilibrium state, we can fully ($100\%$) identify the most influential component index correctly. Even for $50\%$ of random perturbation on the equilibrium state, we can robustly ($87.6\%$) identify the most influential component index correctly. These observations confirm that the proposed VV for the linearized dynamics has an acceptable performance in terms of robustly identifying the most influential component index correctly. Finally, we note that computing VVs and subsequently identifying influential network components would be computationally prohibitive for even mid-scale networks, meaning that the proposed linearization-based approach is an adequate compromise to solve a complex network science problem.

\subsection{Correlation between the Quasi-Exact Variations and VVs}

In this section, we compute the correlation between the quasi-exact variations (i.e., $J$) and VVs (i.e., $v$) for various input types using \eqref{EVs} and Propositions \ref{Propo1}--\ref{Propo3}. To that end, considering the equilibrium initial state ($\bar{q}(0) = 0$), we use the MATLAB built-in function $\texttt{corr}$ with three types: Pearson, Kendall, and Spearman. Tab. \ref{tab:3} illustrates the correlation values corresponding to the $66$th benchmark WDN in the equilibrium initial state ($\bar{q}(0) = 0$). As Tab. \ref{tab:3} depicts, the quasi-exact variation is correlated with VV for all scenarios (particularly, highly correlated for the non-impulse inputs), that is, we effectively identify the network's most and least influential pipes by efficiently computing VVs while avoiding the computationally expensive evaluation of the quasi-exact variations. Moreover, we observe that shifted pulse input with $\mathcal{S}_2[\cdot]$ and shifted impulse input with $\mathcal{S}_3[\cdot]$ respectively attain the best and worst performances in terms of VVs' effectiveness. Such different performances can be justified by the fact that since $\mathrm{sa}(F + F^\top) = -0.0018 < 0$ holds, we get $J_m(\alpha,T) = |\alpha|\|G \mathrm{e}_m^M\|$ for the case of shifted impulse input with $\mathcal{S}_3[\cdot]$ and the equilibrium initial state ($\bar{q}(0) = 0$). Note that $|\alpha|\|G \mathrm{e}_m^M\|$ does not depend on $F$ and $T-\mu$ while the corresponding VV, i.e., $v_m = \mathrm{e}_m^{M \top} G^\top W_{T-\mu}G \mathrm{e}_m^M$ depends on $F$ and $T-\mu$ due to the dependency of $W_{T-\mu}$ through the Lyapunov equation \eqref{g1}. Then, in the equilibrium initial state ($\bar{q}(0) = 0$), it can be argued that, unlike the case of shifted pulse input with $\mathcal{S}_2[\cdot]$, \eqref{EVs} seems to be less informative for the case of shifted impulse input with $\mathcal{S}_3[\cdot]$.

\begin{table}[t]
    \centering
    \caption{Correlation between the quasi-exact variations and VVs corresponding to the $66$th benchmark WDN in the equilibrium initial state ($\bar{q}(0) = 0$).}
    \begin{tabular}{lccc}
        \toprule
        \rowcolor{tableblue} \thead{Input} & \thead{Pearson} & \thead{Kendall} & \thead{Spearman} \\
        \midrule
        Shifted step input, $\mathcal{S}_1[\cdot]$ & $0.9983$ & $0.9920$ & $0.9998$ \\
        Shifted pulse input, $\mathcal{S}_1[\cdot]$ & $1.0000$ & $0.9949$ & $0.9999$ \\
        Shifted pulse input, $\mathcal{S}_2[\cdot]$ & $1.0000$ & $1.0000$ & $1.0000$ \\
        Shifted impulse input, $\mathcal{S}_3[\cdot]$ & $0.8346$ & $0.8634$ & $0.9661$ \\
        \bottomrule
    \end{tabular}
    \label{tab:3}
\end{table}

To showcase the effect of deviation from the equilibrium initial state (i.e., non-equilibrium initial state ($\bar{q}(0) \neq 0$)) on the correlation between the quasi-exact variations and VVs, we consider the following nonzero $\bar{q}(0)$:
\begin{align} \label{qbarnz}
    & \bar{q}(0) = \frac{\varrho \|U_2^\top q^\ast\|}{\|\varphi\|} U_2\varphi,
\end{align}
where $\varrho > 0$ and $\varphi = 2*\texttt{rand}(K,1)-\texttt{ones}(K,1)$. Note that MATLAB built-in functions $\texttt{rand}(K,1)$ and $\texttt{ones}(K,1)$ respectively generate a uniformly distributed random $K$-dimensional vector and the $K$-dimensional vector of all ones. Depending on the input type, we select $\alpha$ as follows:
\begin{align} \label{alfqbnz}
    & \alpha = \begin{cases}
    \frac{\|U_2^\top q^\ast\|}{\underset{m \in \mathbb{N}_M}{\max} \|F^{-1}G\mathrm{e}_m^M\|}, & \mathrm{Shifted~Step/Pulse~Input}\\
    \frac{\|U_2^\top q^\ast\|}{\underset{m \in \mathbb{N}_M}{\max} \|G\mathrm{e}_m^M\|}, & \mathrm{Shifted~Impulse~Input}
\end{cases}.
\end{align}
Considering the $\bar{q}(0)$ in \eqref{qbarnz} and the $\alpha$ in \eqref{alfqbnz} for two values of $\varrho \in \{0.25,0.5\}$, Tab. \ref{tab:4} illustrates the correlation values corresponding to the $66$th benchmark WDN in the non-equilibrium initial state ($\bar{q}(0) \neq 0$). As Tab. \ref{tab:4} depicts, for the non-impulse inputs, the correlation values deteriorate by intensifying the deviation from the equilibrium initial state (i.e., by increasing the value of $\varrho$). As highlighted previously, \eqref{EVs} seems less informative for the impulse input. Comparing the results of Tabs. \ref{tab:3} and \ref{tab:4}, we realize that the proposed VVs are more representative in the equilibrium-like initial state ($\bar{q}(0) \to 0$) for the case of non-impulse inputs.  

\begin{table}[t]
    \centering
    \caption{Correlation between the quasi-exact variations and VVs corresponding to the $66$th benchmark WDN in the non-equilibrium initial state ($\bar{q}(0) \neq 0$) for two values of $\varrho \in \{0.25,0.5\}$. [Top]: $\varrho = 0.25$, [Bottom]: $\varrho = 0.5$.}
    \begin{tabular}{lccc}
        \toprule
        \rowcolor{tableblue} \thead{Input} & \thead{Pearson} & \thead{Kendall} & \thead{Spearman} \\
        \midrule
        \rowcolor{paleblue} \multicolumn{4}{l}{\itshape $\varrho = 0.25$} \\
        Shifted step input, $\mathcal{S}_1[\cdot]$ & $0.8723$ & $0.2864$ & $0.3477$ \\
        Shifted pulse input, $\mathcal{S}_1[\cdot]$ & $0.8424$ & $0.2601$ & $0.3158$ \\
        Shifted pulse input, $\mathcal{S}_2[\cdot]$ & $0.8436$ & $0.2602$ & $0.3158$ \\
        Shifted impulse input, $\mathcal{S}_3[\cdot]$ & $0.9291$ & $0.3177$ & $0.3878$ \\
        \midrule
        \rowcolor{paleblue} \multicolumn{4}{l}{\itshape $\varrho = 0.5$} \\
        Shifted step input, $\mathcal{S}_1[\cdot]$ & $0.8320$ & $0.1842$ & $0.2253$ \\
        Shifted pulse input, $\mathcal{S}_1[\cdot]$ & $0.7393$ & $0.1510$ & $0.1845$ \\
        Shifted pulse input, $\mathcal{S}_2[\cdot]$ & $0.7396$ & $0.1511$ & $0.1845$ \\
        Shifted impulse input, $\mathcal{S}_3[\cdot]$ & $0.9349$ & $0.2026$ & $0.2472$ \\
        \bottomrule
    \end{tabular}
    \label{tab:4}
\end{table}

Furthermore, we include a comparison between the proposed dynamic control theoretic VV and the graph-theoretic edge betweenness centrality (EBC) \cite{freeman1977set}. The normalized EBC can be computed as follows: 
\begin{align*}
    C'_B(e) := \frac{2}{N(N-1)} \sum_{s \neq t \in V} \frac{\sigma(s,t|e)}{\sigma(s,t)},
\end{align*}
where $\sigma(s,t)$ and $\sigma(s,t|e)$ denote the total number of shortest paths between nodes $s$ and $t$, and the number of those paths passing through edge $e$, respectively. To compute the normalized EBC $C'_B$, we apply the brain connectivity toolbox \cite{rubinov2010complex} to the weighted adjacency matrix $A = \mathrm{dg}(\mathrm{dg}(CDC^\top)) - CDC^\top$. Tab. \ref{tab:VVEBC} represents the correlation between the VVs (i.e., $v$) and the normalized EBC (i.e., $C'_B$). Since all the correlations in Tab. \ref{tab:VVEBC} are negative, we conclude that the graph-theoretic EBC is unable to effectively quantify the influence of pipes as it overlooks the dynamics of the WDN. We also realize that the most influential pipe is the $14$th pipe connecting the $15$th node and the $37$th node and the least influential pipe is the $485$th pipe connecting the $223$rd node and the $337$th node, indicating that the identified most and least influential pipes are nonidentical to the ones identified by the VV depicted by Fig. \ref{fig:1} [Left].

\begin{table}[t]
    \centering
    \caption{Correlation between the VVs and the normalized EBC corresponding to the $66$th benchmark WDN.}
    \begin{tabular}{lccc}
        \toprule
        \rowcolor{tableblue} \thead{Input} & \thead{Pearson} & \thead{Kendall} & \thead{Spearman} \\
        \midrule
        Shifted step input, $\mathcal{S}_1[\cdot]$ & $-0.1387$ & $-0.1331$ & $-0.2071$ \\
        Shifted pulse input, $\mathcal{S}_1[\cdot]$ & $-0.1430$ & $-0.1320$ & $-0.2055$ \\
        Shifted pulse input, $\mathcal{S}_2[\cdot]$ & $-0.1426$ & $-0.1318$ & $-0.2051$ \\
        Shifted impulse input, $\mathcal{S}_3[\cdot]$ & $-0.0939$ & $-0.1285$ & $-0.2007$ \\
        \bottomrule
    \end{tabular}
    \label{tab:VVEBC}
\end{table}

\section{Concluding Remarks} \label{Con}

This paper introduces a control-theoretic node centrality-based measure to rank the network components (pipes) based on their influence on the dynamics and topology of the WDN in the case of input changes. This approach can identify the network's most and least influential pipes in WDNs.

Here, we are ready to provide comprehensive answers to \qlab{Q1}--\qlab{Q5} posed in Section \ref{sec:ProFor}:
\begin{itemize}[\IEEEsetlabelwidth{Q5.}\setlength{\itemsep}{3pt}]
    \item[\qlab{Q1.}] Is it possible to incorporate both the dynamics and topology of a WDN to identify its most and least influential pipes?\\[1pt]
    \qlab{A1.} Yes. We show that by appropriately defining a node centrality-based VV, one can simultaneously incorporate the dynamics and topology of a WDN to identify its most and least influential pipes. It is noteworthy that, according to the definition of VV via $v_m$ in \eqref{GamForm}, $\tilde{q}(t;m,\alpha)$ depends on $t$ (dynamics) and $F,G$ in \eqref{LODEs} that depends on the incidence matrix $C$ (topology) via $B$ and $q^{\ast}$. Thus, VV can integrate both the dynamics and topology of a WDN. 
    \item[\qlab{Q2.}] Can one derive mathematical expressions for VV in terms of dynamics and topology of the network without evaluating the exact variations associated with the various inputs?\\[1pt]
    \qlab{A2.} Yes. Propositions \ref{Propo1}--\ref{Propo3} certify that there exist mathematical expressions for VV in terms of the dynamics and topology of the network. Such mathematical expressions facilitate vulnerability analysis without requiring computationally expensive exact evaluation of the variations associated with the various inputs.
    \item[\qlab{Q3.}] Is the VV-based influence of the network pipes dependent on the input types and their timing parameters?\\[1pt]
    \qlab{A3.} Yes. Through extensive numerical simulations visualized by Fig. \ref{fig:3}, we investigate the effects of input types and timing parameters on the VV-based influence of the network pipes. Fig. \ref{figg} corroborates that depending on the specifications of the benchmark WDN, we may observe a totally different parametric dependency and higher sensitivity levels associated with the VV-based influences for different inputs and three timing parameters $(T,\mu,\eta)$.
    \item[\qlab{Q4.}] Is there any graph-theoretic interpretation for the VV-based network's insignificant components (pipes and nodes)?\\[1pt]
    \qlab{A4.} Yes. Interestingly, we discover a graph-theoretic interpretation for the VV-based network's insignificant components (pipes and nodes). Specifically, in Section \ref{SecD}, we propose a graph-theoretic iterative procedure to characterize the subset of the VV-based network's insignificant components. Such characterization enables us to derive a lower bound for the total number of insignificant components.
    \item[\qlab{Q5.}] Can one induce a centrality measure from VV to additionally identify the network's most and least influential nodes in WDNs?\\[1pt]
    \qlab{A5.} Yes. We can define a centrality measure based on the proposed VV to identify the network's most and least influential nodes in WDNs. Such a centrality measure can be interpreted as the dynamic version of the \textit{degree}, a static graph-theoretic centrality measure \cite{freeman1977set}.
\end{itemize}

\parhead{Limitations and Future Directions.} A potential future direction could be extending the centrality measures to a wider class of more complicated WDNs, incorporating the effects of pumps in the modeling. As another pertinent research direction, one could consider the development of systematic centrality-based maintenance (optimization and design) of WDNs, leading to insightful practical guidelines for water engineers. We have detailed how to compute VV for three main input types. Computing VV for the sinusoidal disturbance inputs can be considered as another interesting future direction. As a limitation of the current study, one requires some trial and error to tune some parameters. Therefore, outlining a procedure to tune parameters, e.g., the time horizon length parameter $T$ and a subsequent theoretical analysis would be very beneficial as a future direction. As another limitation of the current study, the proposed VV in the current study has been specialized to the special NDAE dynamics borrowed from the literature. Then, another future work could be searching for simpler yet representative surrogate models to compute the VV using a wide range of perturbations for the initial conditions, and hence providing a more generic VV computation than the one we present in the current study.

\begin{appendices}

\section{Construction of the NDAE Representation of WDNs} \label{CNDAE}

The dynamical equation representing the transient flows through the pipes connecting the nodes can compactly be expressed as
\begin{align} \label{TFTP}
    & \dot{q}(t) = D[C^\top h(t) - r(q(t)) \odot q(t) \odot |q(t)| - u(t)],
\end{align}
with $C = \begin{bmatrix}
    C_v^\top & C_f^\top
\end{bmatrix}^\top$ and $h(t) = \begin{bmatrix}
    h_v(t)^\top & h_f^\top
\end{bmatrix}^\top$, and applying the conservation of water mass (i.e., Kirchhoff's current law) to the nodes with varying total heads, we get
\begin{align} \label{CWM}
    & C_v q(t) + \kappa^{\mathrm{ext}} = 0.
\end{align}

Taking the derivative with respect to $t$ from \eqref{CWM}, we get $C_v \dot{q}(t) = 0$. By substituting \eqref{TFTP} in $C_v \dot{q}(t) = 0$ and noting that $C_v D C_v^\top$ is invertible (due to the connectivity of the corresponding sub-graph), we get
\begin{subequations} \label{hveq}
    \begin{align}
    & h_v(t) = (C_v D C_v^\top)^{-1} C_v D [p(q(t))-C_f^\top h_f + u(t)],\\
    & p(q(t)) = r(q(t)) \odot q(t) \odot |q(t)|.
    \end{align}
\end{subequations}

By substituting \eqref{hveq} in \eqref{TFTP} and defining $B := B(C_v,D) = DC_v^\top (C_vDC_v^\top)^{-1}C_vD-D$, we have
\begin{align*}
    \dot{q}(t) &= D[C_v^\top h_v(t) + C_f h_f - p(q(t)) - u(t)]\\
    &= B [p(q(t))-C_f^\top h_f + u(t)]. 
\end{align*}Thus, we obtain NDAE representation \eqref{NDAEs}.

\section{Proof of Proposition \ref{Propo1}} \label{PF1}

According to \eqref{GamForm}, \eqref{pSSI}, and \eqref{Sch}, for the shifted step input $\bar{u}(t) = \alpha \mathrm{1}_{\mathbb{R}_+}(t-\mu) \mathrm{e}_m^M$, we have
\begin{align*}
    v_m & = \sqrt{\frac{1}{T} \int_0^T \bigg \|\frac{\partial \tilde{q}(t;m,\alpha)}{\partial \alpha} \bigg \|^2 dt}\\
    & = \sqrt{\frac{1}{T} \int_0^T \bigg (\frac{\partial \tilde{q}(t;m,\alpha)}{\partial \alpha} \bigg )^\top \frac{\partial \tilde{q}(t;m,\alpha)}{\partial \alpha}  dt}\\
    & = \sqrt{\frac{1}{T} \int_\mu^T ((e^{F(t-\mu)}-I_K)p_m^K)^\top (e^{F(t-\mu)}-I_K)p_m^K dt}\\
    & = \sqrt{\frac{1}{T} \int_0^{T-\mu} [(e^{F\psi}-I_K)p_m^K]^\top (e^{F\psi}-I_K)p_m^K d\psi}\\
    & = \sqrt{\frac{1}{T} p_m^{K \top} \int_0^{T-\mu} e^{F^\top \psi} e^{F \psi}-e^{F^\top \psi}-e^{F \psi}+I_K d\psi p_m^K}\\
    & = \sqrt{\frac{1}{T} p_m^{K \top} (\chi_1 - \chi_2 - \chi_3 + \chi_4) p_m^K},\\
    & \chi_1 := \int_0^{T-\mu} e^{F^\top \psi} e^{F \psi} d \psi, \chi_2 := \int_0^{T-\mu} e^{F^\top \psi}d \psi,\\
    & \chi_3 := \int_0^{T-\mu} e^{F \psi}d \psi,\chi_4 := \int_0^{T-\mu} I_K d \psi.
\end{align*}
Then, we have
\begin{align*}
    & \chi_4 = \int_0^{T-\mu} I_K d \psi = (T-\mu) I_K,\\
    & \chi_3 = \int_0^{T-\mu} e^{F \psi}d \psi = F^{-1} e^{F \psi}|_0^{T-\mu} = F^{-1}(e^{F (T-\mu)}-I_K),\\
    & \chi_2 = \chi_3^\top = (e^{F^\top (T-\mu)}-I_K) (F^{-1})^\top.
\end{align*}
Note that the following identity holds for $\chi_1$:
\begin{align*}
    & F^\top \chi_1 + \chi_1 F = \int_0^{T-\mu} \frac{d}{d\psi}(e^{F^\top \psi} e^{F \psi}) d \psi = e^{F^\top \psi} e^{F \psi}|_0^{T-\mu},\\
    & F^\top \chi_1 + \chi_1 F + I_K - e^{F^\top (T-\mu)} e^{F (T-\mu)} = 0.
\end{align*}
Then, $\chi_1$ can be computed by solving a Lyapunov equation. According to \cite{bhatia1997and}, the Lyapunov equation has a unique solution if and only if $F^\top$ and $-F$ do not share any eigenvalue. We highlight that the eigenvalues of $F^\top$ are the same as those of $F$. Since the state matrix $F$ is stable (Hurwitz), then its eigenvalues all lie on the open left half plane and $F^\top$ and $-F$ do not share any eigenvalue, implying that the Lyapunov equation has a unique solution $\chi_1$. For such a unique solution $\chi_1$, we have the following identities \cite{bhatia1997and}:
\begin{align*}
    \chi_1 =&~ \int_0^{\infty} e^{F^\top t}(I_K -e^{F^\top(T-\mu)}e^{F(T-\mu)})e^{F t} dt,\\
    \mathrm{vec}(\chi_1) =&~ (I_K \otimes F^\top + F^\top \otimes I_K)^{-1}\\ 
    & ~\times \mathrm{vec}(e^{F^\top(T-\mu)}e^{F(T-\mu)}-I_K).
\end{align*}

To show that $\chi_1$ is positive-definite, i.e., $\chi_1 \succ 0$ holds, we must prove that $\omega^\top \chi_1 \omega > 0$ holds for all non-zero vectors $\omega \in \mathbb{R}^K$. Then, we have
\begin{align*}
    & \omega^\top \chi_1 \omega = \omega^\top \int_0^{T-\mu} e^{F^\top \psi} e^{F \psi} d \psi \omega\\
    & = \int_0^{T-\mu} \omega^\top e^{F^\top \psi} e^{F \psi} \omega d \psi = \int_0^{T-\mu} (e^{F \psi} \omega)^\top e^{F \psi} \omega d \psi\\
    & = \int_0^{T-\mu} \|e^{F \psi} \omega\|^2 d \psi.
\end{align*}
Now, we prove by contradiction that $\|e^{F \psi}\omega\|^2$ cannot be $0$. Assume that $\|e^{F \psi}\omega\|^2 = 0$, then $e^{F \psi}\omega = 0$. Since $F$ is stable (Hurwitz), $e^{F \psi}$ is invertible and its inverse is $e^{-F \psi}$. By pre-multiplying $e^{F \psi}\omega = 0$ by $e^{-F \psi}$, we get $\omega = 0$ which is impossible. Thus, $\|e^{F \psi}\omega\|^2 > 0$ holds for all non-zero vectors $\omega$ and $\omega^\top \chi_1 \omega = \int_0^{T-\mu} \|e^{F \psi} \omega\|^2 d \psi > 0$ holds for all non-zero vectors $\omega$. Thus, $\chi_1$ is the unique positive-definite solution of the Lyapunov equation. Thus, substituting the computed $\chi_1$, $\chi_2$, $\chi_3$, and $\chi_4$ in $\sqrt{\frac{1}{T} p_m^{K \top} (\chi_1 - \chi_2 - \chi_3 + \chi_4) p_m^K}$ and noting that $p_m^K := F^{-1}G \mathrm{e}_m^M$ holds, completes the proof.

\section{Proof of Proposition \ref{Propo2}} \label{PF2}
According to \eqref{GamForm}, \eqref{pSPI}, and \eqref{Sch}, for the shifted pulse input $\bar{u}(t) = \alpha[\mathrm{1}_{\mathbb{R}_+}(t-\mu)-\mathrm{1}_{\mathbb{R}_+}(t-\mu-\eta)] \mathrm{e}_m^M$, we have
\begin{align*}
    & v_m = \sqrt{\frac{1}{T} \int_0^T \bigg \|\frac{\partial \tilde{q}(t;m,\alpha)}{\partial \alpha} \bigg \|^2 dt}\\
    & = \sqrt{\frac{1}{T} \int_0^T \bigg (\frac{\partial \tilde{q}(t;m,\alpha)}{\partial \alpha} \bigg )^\top \frac{\partial \tilde{q}(t;m,\alpha)}{\partial \alpha} dt} = \sqrt{\frac{1}{T} (\beta_1 + \beta_2)}\\
    & \beta_1 := \int_\mu^\xi [(e^{F(t-\mu)}-I_K)p_m^K]^\top (e^{F(t-\mu)}-I_K)p_m^K dt,\\
    & \beta_2 := \int_\xi^T [(e^{F\eta}-I_K)p_m^K]^\top (e^{F\eta}-I_K)p_m^K dt.
\end{align*}
We compute $\beta_1$ and $\beta_2$, respectively.

For $\beta_1$, we have
\begin{align*}
    \beta_1 & = \int_\mu^\xi [(e^{F(t-\mu)}-I_K)p_m^K]^\top (e^{F(t-\mu)}-I_K)p_m^K dt\\
    & = \int_0^{\eta} [(e^{F\psi}-I_K)p_m^K]^\top (e^{F\psi}-I_K)p_m^K d\psi\\
    & = p_m^{K \top} \int_0^{\eta} e^{F^\top \psi} e^{F \psi}-e^{F^\top \psi}-e^{F \psi}+I_K d\psi p_m^K\\
    & = p_m^{K \top} (\chi_1 - \chi_2 - \chi_3 + \chi_4) p_m^K,\\
    & \chi_1 := \int_0^{\eta} e^{F^\top \psi} e^{F \psi} d \psi, \chi_2 := \int_0^{\eta} e^{F^\top \psi}d \psi,\\
    & \chi_3 := \int_0^{\eta} e^{F \psi}d \psi,\chi_4 := \int_0^{\eta} I_K d \psi.
\end{align*}
Then, we have
\begin{align*}
    & \chi_4 = \int_0^{\eta} I_K d \psi = \eta I_K,\\
    & \chi_3 = \int_0^{\eta} e^{F \psi}d \psi = F^{-1} e^{F \psi}|_0^{\eta} = F^{-1}(e^{F \eta}-I_K),\\
    & \chi_2 = \chi_3^\top = (e^{F^\top \eta}-I_K) (F^{-1})^\top.
\end{align*}
Note that the following identity holds for $\chi_1$:
\begin{align*}
    & F^\top \chi_1 + \chi_1 F = \int_0^{\eta} \frac{d}{d\psi}(e^{F^\top \psi} e^{F \psi}) d \psi = e^{F^\top \psi} e^{F \psi}|_0^{\eta},\\
    & F^\top \chi_1 + \chi_1 F + I_K - e^{F^\top \eta} e^{F \eta} = 0.
\end{align*}
Then, $\chi_1$ can be computed by solving a Lyapunov equation. Similarly, it can be shown that $\chi_1$ is the unique positive-definite solution of the Lyapunov equation. For such a unique solution $\chi_1$, we have the following identities \cite{bhatia1997and}:
\begin{align*}
    \chi_1 &= \int_0^{\infty} e^{F^\top t}(I_K -e^{F^\top \eta}e^{F \eta})e^{F t} dt,\\
    \mathrm{vec}(\chi_1) &= (I_K \otimes F^\top + F^\top \otimes I_K)^{-1} \mathrm{vec}(e^{F^\top \eta}e^{F \eta}-I_K).
\end{align*} 
Substituting $\chi_1$, $\chi_2$, $\chi_3$, and $\chi_4$ in $p_m^{K \top} (\chi_1 - \chi_2 - \chi_3 + \chi_4) p_m^K$, $\beta_1$ can be computed.

For $\beta_2$, we have
\begin{align*}
    \beta_2 &= \int_\xi^T [(e^{F\eta}-I_K)p_m^K]^\top (e^{F\eta}-I_K)p_m^K dt\\
    &= (T-\mu-\eta) p_m^{K \top} (e^{F^\top \eta}-I_K)(e^{F \eta}-I_K) p_m^K.
\end{align*}Thus, substituting the computed $\beta_1$ and $\beta_2$ in $\sqrt{\frac{1}{T} (\beta_1 + \beta_2)}$ and noting that $p_m^K := F^{-1}G \mathrm{e}_m^M$ holds, completes the proof.

\section{Proof of Proposition \ref{Propo3}} \label{PF3}

According to \eqref{GamForm}, \eqref{pSPI}, and \eqref{Schh}, for the shifted pulse input $\bar{u}(t) = \alpha[\mathrm{1}_{\mathbb{R}_+}(t-\mu)-\mathrm{1}_{\mathbb{R}_+}(t-\mu-\eta)] \mathrm{e}_m^M$, we have
\begin{align*}
    v_m &= \max \Bigg \{\sqrt{\frac{1}{\eta}\int_{\mu}^{\xi} \bigg \|\frac{\partial \tilde{q}(t;m,\alpha)}{\partial \alpha} \bigg \|^2 dt},\bigg \|\frac{\partial \tilde{q}(\xi;m,\alpha)}{\partial \alpha} \bigg \| \Bigg \}.
\end{align*}
We compute $\int_{\mu}^{\xi} \big \|\frac{\partial \tilde{q}(t;m,\alpha)}{\partial \alpha} \big \|^2 dt$ and $\big \|\frac{\partial \tilde{q}(\xi;m,\alpha)}{\partial \alpha} \big \|$, respectively.

For $\int_{\mu}^{\xi} \big \|\frac{\partial \tilde{q}(t;m,\alpha)}{\partial \alpha} \big \|^2 dt$, we have
\begin{align*}
    & \int_{\mu}^{\xi} \bigg \|\frac{\partial \tilde{q}(t;m,\alpha)}{\partial \alpha} \bigg \|^2 dt = \int_{\mu}^{\xi} \bigg (\frac{\partial \tilde{q}(t;m,\alpha)}{\partial \alpha} \bigg )^\top \frac{\partial \tilde{q}(t;m,\alpha)}{\partial \alpha} dt\\
    & = \int_\mu^\xi [(e^{F(t-\mu)}-I_K)p_m^K]^\top (e^{F(t-\mu)}-I_K)p_m^K dt \\
    & = \int_0^{\eta} [(e^{F\psi}-I_K)p_m^K]^\top (e^{F\psi}-I_K)p_m^K d\psi \\
    & = p_m^{K \top} (\chi_1 - \chi_2 - \chi_3 + \chi_4) p_m^K,
\end{align*}
where $\chi_4 = \eta I_K$, $\chi_3 = F^{-1}(e^{F \eta}-I_K)$, $\chi_2 = \chi_3^\top = (e^{F^\top \eta}-I_K) (F^{-1})^\top$, and $\chi_1$ denotes the unique positive-definite solution of the following Lyapunov equation:
\begin{align*}
    & F^\top \chi_1 + \chi_1 F + I_K - e^{F^\top \eta} e^{F \eta} = 0.
\end{align*}

For $\big \|\frac{\partial \tilde{q}(\xi;m,\alpha)}{\partial \alpha} \big \|$, we have
\begin{align*}
    & \bigg \|\frac{\partial \tilde{q}(\xi;m,\alpha)}{\partial \alpha} \bigg \| \\
    & = \|(e^{F\eta}-I_K)p_m^K\| = \sqrt{[(e^{F\eta}-I_K)p_m^K]^\top (e^{F\eta}-I_K)p_m^K}\\
    & = \sqrt{p_m^{K \top} (e^{F^\top \eta}-I_K)(e^{F \eta}-I_K) p_m^K}.
\end{align*}

Thus, substituting the computed $\int_{\mu}^{\xi} \big \|\frac{\partial \tilde{q}(t;m,\alpha)}{\partial \alpha} \big \|^2 dt$ and $\big \|\frac{\partial \tilde{q}(\xi;m,\alpha)}{\partial \alpha} \big \|$ in $\max \Big \{\sqrt{\frac{1}{\eta}\int_{\mu}^{\xi} \big \|\frac{\partial \tilde{q}(t;m,\alpha)}{\partial \alpha} \big \|^2 dt},\big \|\frac{\partial \tilde{q}(\xi;m,\alpha)}{\partial \alpha} \big \| \Big \}$ and noting that $p_m^K := F^{-1}G \mathrm{e}_m^M$ holds, completes the proof.

\section{Proof of Proposition \ref{Propo4}} \label{PF4}

Note that $C_v B = 0$ holds. Since $B$ is a symmetric matrix, i.e., $B = B^\top$ holds, we get $B C_v^\top = B^\top C_v^\top = (C_v B)^\top = 0$ and equivalently $B C_v(i,:)^\top = 0$ holds for all $i \in \{1,\dots,N_v\}$. 
    
    \textit{(i)} Observe that if the $i$-th row of $C_v$ has only one non-zero element (i.e., the $i$-th node with a varying total head is degree-$1$), then $C_v(i,:) = \mathrm{e}_{m_i}^{M \top}$ or $C_v(i,:) = -\mathrm{e}_{m_i}^{M \top}$ holds and consequently $B \mathrm{e}_{m_i}^M = B C_v(i,:)^\top = 0$ or $B \mathrm{e}_{m_i}^M = -B C_v(i,:)^\top = 0$ holds meaning that the $m_i$-th pipe satisfies $B \mathrm{e}_m^M = 0$ for $m = m_i$. 

    \textit{(ii)} Observe that if the $i$-th and $j$-th rows of $C_v$ have one and two non-zero elements, respectively (i.e., the $i$-th and $j$-th nodes with varying total heads are degree-$1$ and degree-$2$, respectively) and they are adjacent, then either $C_v(i,:) + C_v(j,:) = \mathrm{e}_{m_j}^{M \top}$ or $C_v(i,:) + C_v(j,:) = -\mathrm{e}_{m_j}^{M \top}$ holds for $m_j \neq m_i$ and consequently either $B \mathrm{e}_{m_j}^M = B (C_v(i,:)^\top + C_v(j,:)^\top) = B C_v(i,:)^\top + B C_v(j,:)^\top = 0 + 0 = 0$ or $B \mathrm{e}_{m_j}^M = -B (C_v(i,:)^\top + C_v(j,:)^\top) = -B C_v(i,:)^\top -B C_v(j,:)^\top = -0-0 = 0$ holds accordingly meaning that in addition to the $m_i$-th pipe, the $m_j$-th pipe also satisfies $B \mathrm{e}_m^M = 0$ for $m = m_j$.

\section{VV-based Network's Most and Least Influential Nodes} \label{SecEConvertedtoAppendix}

In addition to identifying the network's most and least influential pipes, the proposed VV can be tailored to identify the network's most and least influential nodes (e.g., junctions) in WDNs. In the following, we detail the idea enabling such identification.

We highlight that, built upon the proposed VV notion, we can define the following measure to identify the network's most and least influential nodes (e.g., junctions) in WDNs:
\begin{align} \label{NVV}
    w_i &:= \sum_{j \in \mathcal{N}_i} v_{m_j},~ \forall i \in \mathbb{N}_N, 
\end{align}
where $\mathcal{N}_i$ represents the neighboring nodes adjacent to the $i$-th node. Similarly, by sorting the elements of vector $w \in \mathbb{R}^N$, one can identify the network's most and least influential nodes in WDNs. Considering $w_{i}$ defined by \eqref{NVV}, the influence of the $i$-th node can be captured via $w_i$, and the network's most and least influential nodes can be determined as $i^{\mathrm{Most~Influential}} := \argmax_{i \in \mathbb{N}_N} w_i$ and $i^{\mathrm{Least~Influential}} := \argmin_{i \in \mathbb{N}_N} w_i$, respectively. Interestingly, the centrality measure defined by \eqref{NVV} can be interpreted as the dynamic version of the static graph-theoretic centrality measure, i.e., \textit{degree} \cite{freeman1977set}. Note that the presented procedure by Section \ref{SecD} induces a similar notion of insignificant components (nodes) (i.e., $w_i = 0$). Likewise, with a bit of abuse of notation, it provides a lower bound $\underline{n}^{\mathrm{IC}}$ for the total number of insignificant components (nodes) $n^{\mathrm{IC}}$.

\begin{figure}[!t]
    \centering
    \includegraphics[width=\columnwidth]{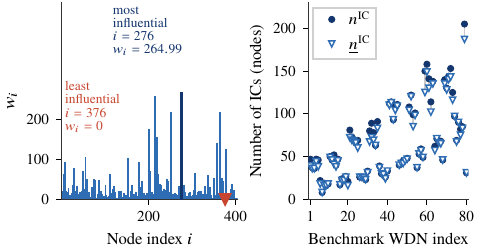}
\caption{[Left]: Distribution of the VV-based node influence vector $w$ across all nodes for the $66$th benchmark WDN. Each bar corresponds to a node, indexed arbitrarily. The highlighted bar corresponds to the most influential node and the marker on the horizontal axis corresponds to one representative least influential (particularly, insignificant) node. [Right]: Total number of insignificant components (nodes) $n^{\mathrm{IC}}$ and the lower bound $\underline{n}^{\mathrm{IC}}$ across all $80$ benchmark WDNs.}
    \label{fig:MILINVV}
\end{figure}

According to the centrality measure \eqref{NVV}, we realize that the network's most influential node and least influential node (in this case, identical to the insignificant node) are the $276$th and the $376$th nodes, respectively, for the $66$th benchmark WDN. Fig. \ref{fig:MILINVV} [Left] depicts the elements of the VV-based vector $w_i$ versus node index $i$ along with the highlighted network's most and least influential nodes for the $66$th benchmark WDN. 

\begin{figure}[!t]
    \centering
    \includegraphics[width=\columnwidth]{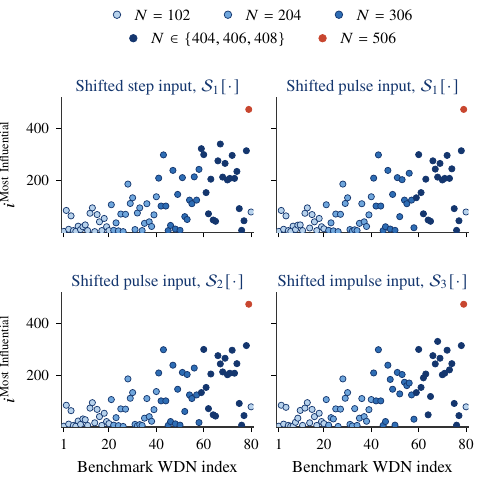}
\caption{VV-based most influential node metric $i^{\mathrm{Most~Influential}}$ for all $80$ benchmark WDNs. [Top Left]: shifted step input with $\mathcal{S}_1[\cdot]$, [Top Right]: shifted pulse input with $\mathcal{S}_1[\cdot]$, [Bottom Left]: shifted pulse input with $\mathcal{S}_2[\cdot]$, [Bottom Right]: shifted impulse input with $\mathcal{S}_3[\cdot]$.}
    \label{fig:NMInds}
\end{figure}

We highlight that in Fig. \ref{fig:MILINVV} [Left], since the least influential node is also an insignificant component in the sense of the VV-based centrality measure \eqref{NVV}, i.e., $w_{376} = 0$ holds, it could be non-unique, which is the case for the $66$th benchmark WDN according to the iterative procedure presented by Section \ref{SecD} as $1 < \underline{n}^{\mathrm{IC}} = 66 \le n^{\mathrm{IC}} = 66$ holds. Then, we emphasize that we have illustrated only one of the least influential nodes. Fig. \ref{fig:MILINVV} [Right] visualizes the total number of insignificant components (nodes) $n^{\mathrm{IC}}$ and the lower bound $\underline{n}^{\mathrm{IC}}$ for all $80$ benchmark WDNs. We visualize the VV-based most influential node metric $i^{\mathrm{Most~Influential}}$ for all $80$ benchmark WDNs in Fig. \ref{fig:NMInds}.

\FloatBarrier

\section{Supplementary Figures for the Parametric Dependency of the VV-based Influences} \label{Suppendix}

\begin{figure}[!t]
    \centering
    \includegraphics[width=\columnwidth]{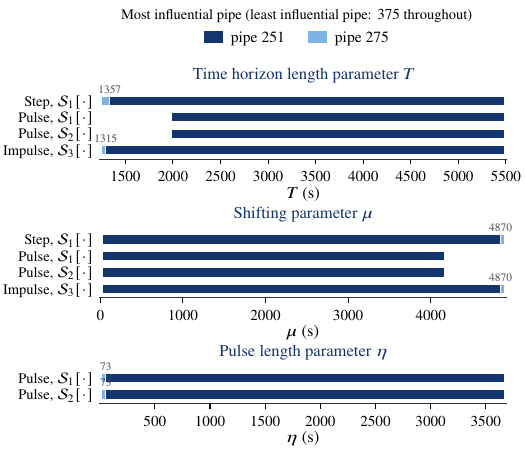}
\caption{Most influential pipe of the $66$th benchmark WDN as a function of the timing parameters. Each strip shows, for the indicated input type and computational operator, the index of the most influential pipe (color coded) over the tested range of the parameter, and the small numbers mark the parameter values at which the most influential pipe changes. [Top]: time horizon length parameter $T$, [Middle]: shifting parameter $\mu$, [Bottom]: pulse length parameter $\eta$. The least influential pipe is the $375$th pipe for all parameter values, input types, and computational operators.}
    \label{fig:3}
\end{figure}

\begin{figure}[!t]
\centering
    \includegraphics[width=\columnwidth]{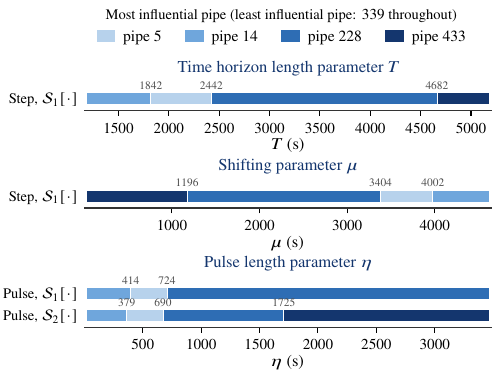}
\caption{Most influential pipe of the $62$nd benchmark WDN as a function of the timing parameters. Each strip shows, for the indicated input type and computational operator, the index of the most influential pipe (color coded) over the tested range of the parameter, and the small numbers mark the parameter values at which the most influential pipe changes. [Top]: time horizon length parameter $T$, shifted step input with $\mathcal{S}_1[\cdot]$, [Middle]: shifting parameter $\mu$, shifted step input with $\mathcal{S}_1[\cdot]$, [Bottom]: pulse length parameter $\eta$, shifted pulse input with $\mathcal{S}_1[\cdot]$ and $\mathcal{S}_2[\cdot]$. The least influential pipe is the $339$th pipe for all parameter values, input types, and computational operators.} \label{figg}
\end{figure}

\begin{figure}[!t]
\centering
\includegraphics[width=\columnwidth]{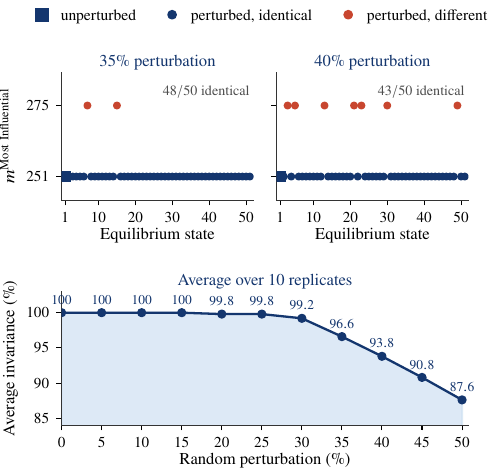}
\caption{Identified most influential components of the $66$th benchmark WDN for different levels of random perturbation; [Top Left]: $35\%$ perturbation, [Top Right]: $40\%$ perturbation, [Bottom]: Average invariance percentage versus random perturbation percentage for the $66$th benchmark WDN.}
    \label{figUP}
\end{figure}

This appendix collects Figs. \ref{fig:3}, \ref{figg}, and \ref{figUP}, which are referred to in Section \ref{NuSim}.

\end{appendices}

\bibliographystyle{IEEEtran}
\bibliography{References}

\makeatletter
\def\@IEEEBIOskipN{1.0\baselineskip}
\expandafter\patchcmd\csname\string\IEEEbiography\endcsname{plus 1fil minus 0\baselineskip}{}{\typeout{BIOPATCH OK}}{\typeout{BIOPATCH FAILED}}
\makeatother
\begin{IEEEbiography}[{\includegraphics[width=1in,height=1.25in,clip,keepaspectratio]{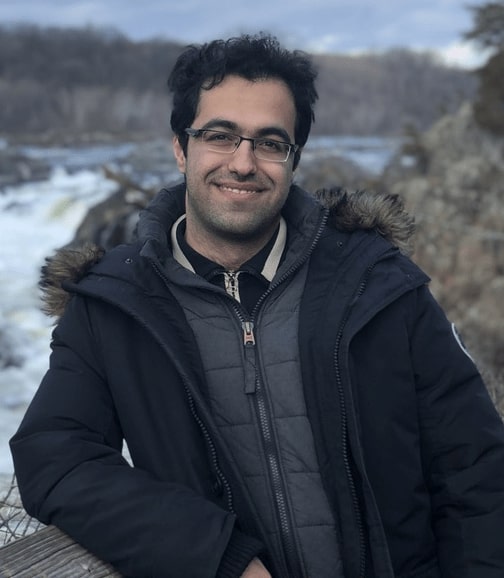}}] {MirSaleh Bahavarnia} received a B.Sc. degree in Electrical Engineering (Control) and a certificate of the minor program in Mathematics from the Sharif University of Technology, Tehran, Tehran, Iran, in 2013 and a Ph.D. degree in Mechanical Engineering (Control) from Lehigh University, Bethlehem, PA, USA, in 2018. He was a Postdoctoral Research Associate with the Department of Electrical and Computer Engineering and the Institute for Systems Research (ISR), University of Maryland, College Park, MD, USA, from 2018 to 2020. He was a Postdoctoral Research Scholar with the Department of Civil and Environmental Engineering, Vanderbilt University, Nashville, TN, USA, from 2022 to 2025. Since 2026, he has been a research scientist with the Department of Civil and Environmental Engineering, Vanderbilt University, Nashville, TN, USA. His research interests include distributed control, feedback control, power systems control, process control, robust control, and traffic control.
\end{IEEEbiography}

\begin{IEEEbiography}[{\includegraphics[width=1in,height=1.25in,clip,keepaspectratio]{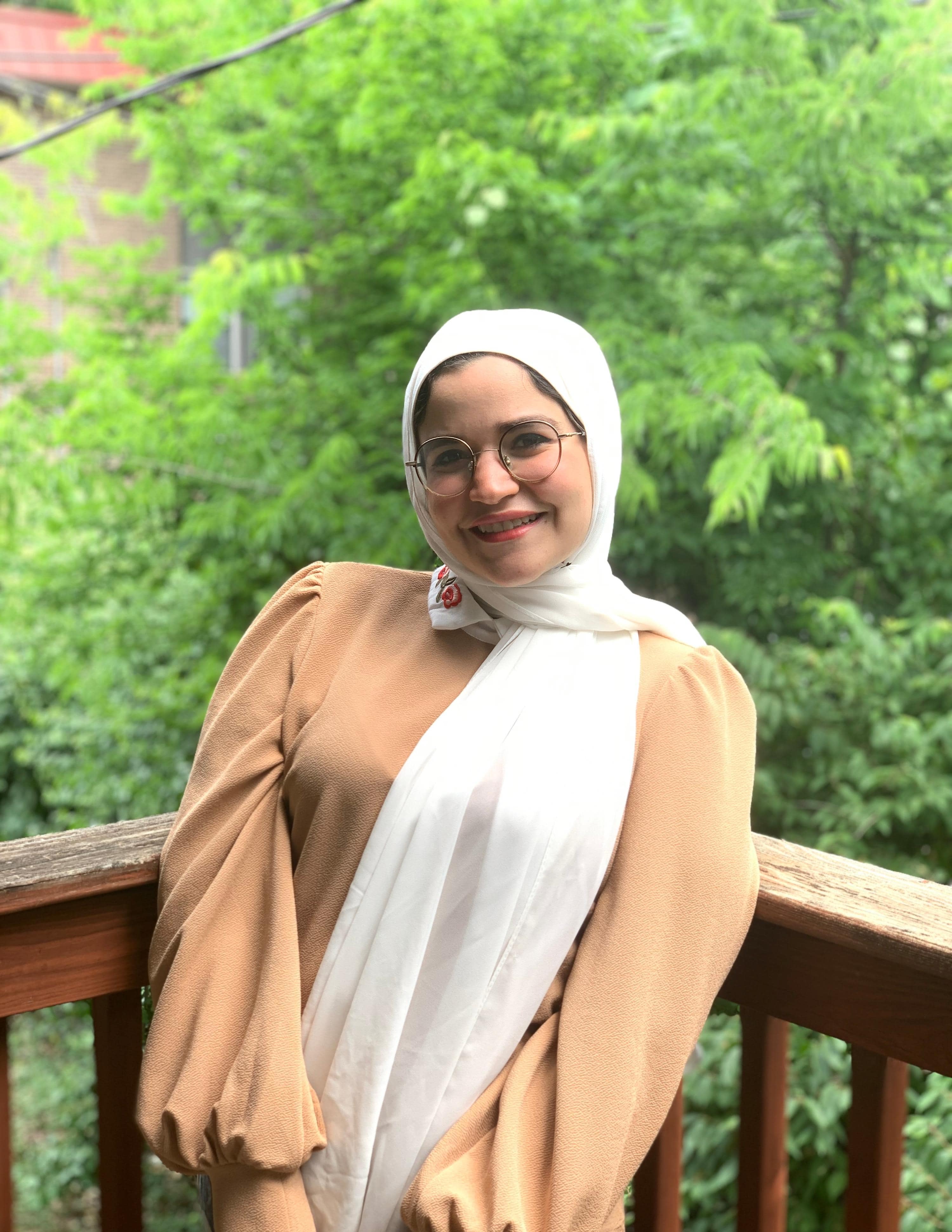}}] {Salma M. Elsherif} received a Ph.D. degree in Civil Engineering from Vanderbilt University, Nashville, TN, USA, in 2025. She holds an M.Sc. in Irrigation and Hydraulic Engineering and a B.Sc. in Civil Engineering from Cairo University (Egypt, 2020, 2016). Her research applies control theory principles to dynamical systems, with a particular focus on civil and environmental infrastructure systems. Specifically, her work focuses on hydraulics, water quality, hydrology, and climate.
\end{IEEEbiography}

\begin{IEEEbiography}[{\includegraphics[width=1in,height=1.25in,clip,keepaspectratio]{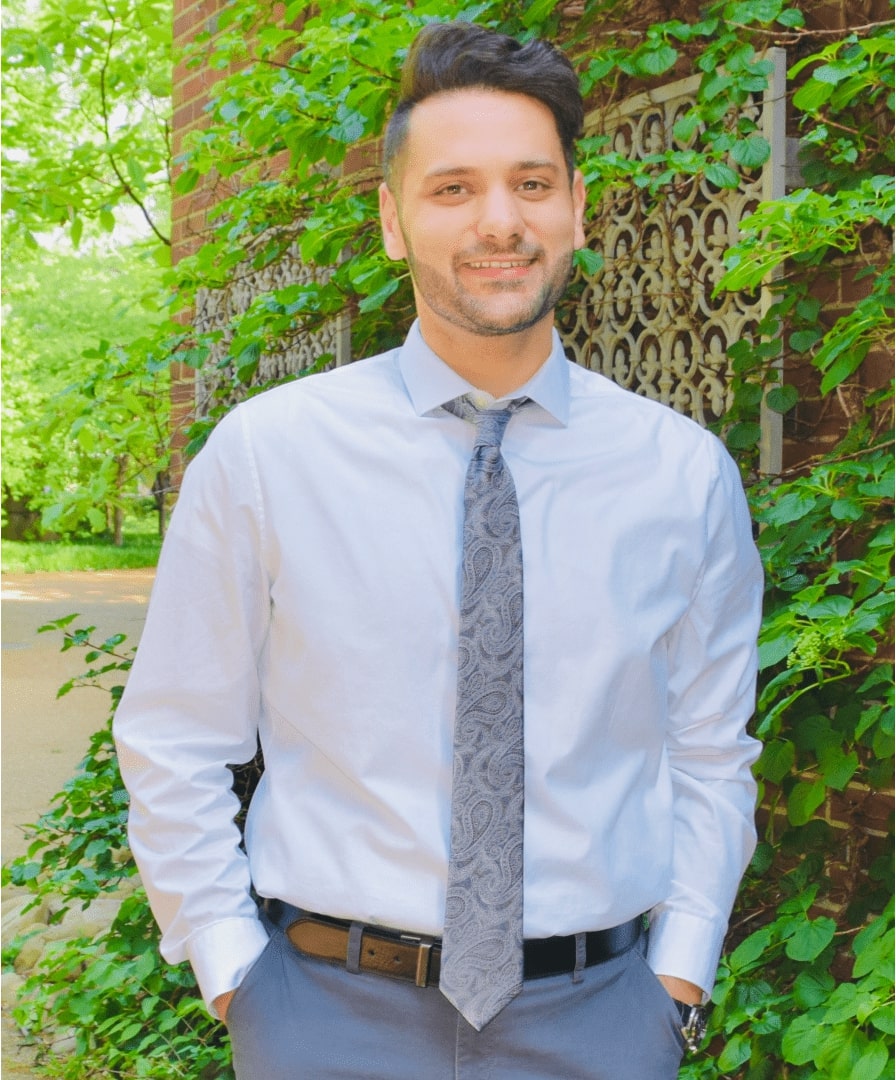}}] {Ahmad F. Taha} is an associate professor with the Department of Civil and Environmental Engineering at Vanderbilt University in Nashville, Tennessee. He has a secondary appointment in Electrical and Computer Engineering. He received his B.E. and Ph.D. degrees in Electrical and Computer Engineering from the American University of Beirut, Lebanon, in 2011 and Purdue University, West Lafayette, IN, USA, in 2015. Before joining Vanderbilt University, Dr. Taha was an assistant professor with the ECE department at the University of Texas at San Antonio (UTSA). Dr. Taha is interested in understanding how complex cyber-physical, urban infrastructures operate, behave, and occasionally \textit{misbehave}. His research focus includes optimization, control, monitoring, and security of infrastructure with power, water, and transportation systems applications. Dr. Taha is an associate editor of \textsc{IEEE Transactions on Control of Network Systems}. 
\end{IEEEbiography}
\vspace*{0pt plus 1fill}

\end{document}

%% file: preamble.tex
\usepackage{color,amsmath,cite}
\usepackage{graphicx}

\IEEEoverridecommandlockouts

\usepackage[table]{xcolor}
\definecolor{navy}{HTML}{14366E}
\definecolor{midblue}{HTML}{2E6DB4}
\definecolor{skyblue}{HTML}{7FB2E5}
\definecolor{paleblue}{HTML}{E8EFF8}
\definecolor{tableblue}{HTML}{D6E4F4}
\definecolor{rulegray}{HTML}{8A8A8A}

\usepackage[linesnumbered,commentsnumbered,ruled,vlined,longend]{algorithm2e}

\usepackage{amsthm} 
\usepackage{amsmath}    
\IEEEoverridecommandlockouts
\usepackage{bm}
\usepackage{amssymb}
\usepackage{url}
\usepackage{booktabs}
\usepackage{empheq}
\newcommand*\widefbox[1]{\fcolorbox{navy}{paleblue}{\hspace{0.25em}#1\hspace{0.25em}}}

\usepackage{tikz}
\usetikzlibrary{positioning, arrows.meta, calc}

\DeclareMathOperator*{\argmax}{arg\ max}
\DeclareMathOperator*{\argmin}{arg\ min}
\makeatother
\DeclareMathAlphabet\mathbfcal{OMS}{cmsy}{b}{n}

\newtheoremstyle{navythm}
  {6pt}{6pt}
  {\itshape}
  {}
  {\color{navy}\bfseries}
  {.}
  { }
  {}
\theoremstyle{navythm}

\newtheorem{mypbm}{Problem}

\newtheorem{myprs}{Proposition}



\allowdisplaybreaks[4]
\usepackage{microtype}
\usepackage{placeins}
\usepackage{etoolbox}

\usepackage[colorlinks = true,
linkcolor = navy,
urlcolor  = navy,
citecolor = navy,
anchorcolor = navy]{hyperref}

\usepackage[noabbrev]{cleveref}

\usepackage{mathtools}

\DeclarePairedDelimiter\abs{\lvert}{\rvert}%
\DeclarePairedDelimiter\norm{\lVert}{\rVert}%

\makeatletter
\let\oldabs\abs
\def\abs{\@ifstar{\oldabs}{\oldabs*}}
\let\oldnorm\norm
\def\norm{\@ifstar{\oldnorm}{\oldnorm*}}
\makeatother

\usepackage[english]{babel}
\usepackage[utf8]{inputenc}
\usepackage[super]{nth}

\usepackage{array}

\usepackage{pifont}
\newcommand{\xmark}{\ding{55}}%

\makeatletter
\def\section{\@startsection{section}{1}{\z@}{3.0ex plus 1.5ex minus 1.5ex}%
{0.7ex plus 1ex minus 0ex}{\normalfont\normalsize\centering\scshape\color{navy}}}%
\def\subsection{\@startsection{subsection}{2}{\z@}{3.5ex plus 1.5ex minus 1.5ex}%
{0.7ex plus .5ex minus 0ex}{\normalfont\normalsize\itshape\color{navy}}}%
\def\subsubsection{\@startsection{subsubsection}{3}{\parindent}{0ex plus 0.1ex minus 0.1ex}%
{0ex}{\normalfont\normalsize\itshape\color{navy}}}%
\makeatother

\newcommand{\parhead}[1]{\noindent\textbf{\textsc{#1}}}
\newcommand{\qlab}[1]{\textcolor{navy}{\textbf{#1}}}

\newcommand{\thead}[1]{\textbf{#1}}

\SetAlFnt{\small}
\SetAlCapFnt{\small\bfseries}
\SetAlCapNameFnt{\small\bfseries\color{navy}}
\SetAlgoCaptionSeparator{:}

\SetKwSty{algkwsty}

\SetNlSty{algnlsty}{}{}

\SetCommentSty{algcommentsty}

\SetKwInput{KwInput}{Input}
\SetKwInput{KwOutput}{Output}
\SetAlgoInsideSkip{smallskip}
\SetAlgoSkip{smallskip}

\usepackage{tcolorbox}
\newtcolorbox{problembox}{colback=paleblue, colframe=navy, boxrule=0.5pt, arc=1pt,
  left=4pt, right=4pt, top=2pt, bottom=2pt, boxsep=1pt}